\pdfoutput=1
\documentclass[epsf,aps,preprint,showpacs,preprintnumbers,amsmath,amssymb,footinbib]{revtex4}
\usepackage{bm}
\usepackage[pdftex]{graphicx}
\usepackage[colorlinks=true, pdfstartview=FitV, linkcolor=red, citecolor=blue, urlcolor=blue]{hyperref}
\newcommand{\beq}{\begin{equation}}
\newcommand{\eeq}{\end{equation}}
\def\acl{A^\text{cl}}

\def\trace{{\rm tr}}
\def\Tc{T_\text{c}}
\def\Nc{N}
\def\Nf{N_\text{f}}

\def\SUN{\text{SU}(\Nc)}
\def\ZN{\text{Z}(\Nc)}
\def\U1{\text{U}(1)}
\def\proj{{\cal P}}
\def\HTLapprox{\stackrel{\text{HTL}}{\approx}}
\newcommand{\Slash}[1]{\ooalign{\hfil/\hfil\crcr$#1$}}
\newcommand{\otherSlash}{\not \!\!}
\newcommand{\TInt}{\int\frac{d^4K}{(2\pi)^4}}
\newcommand{\TTInt}{T \sum_{n=-\infty}^{+\infty} \int\frac{d^3k}{(2\pi)^3}}
\newcommand{\aptl}{\widetilde{p}^{\,a}}
\newcommand{\cktl}{\widetilde{k}^c}
\newcommand{\aPtl}{\widetilde{P}^a}
\newcommand{\bPtl}{\widetilde{P}^b}

\newcommand{\cKtl}{\widetilde{K}^c}

\newcommand{\eKtl}{\widetilde{K}^e}

\newcommand{\aQtl}{\widetilde{Q}^a}
\newcommand{\bQtl}{\widetilde{Q}^b}
\newcommand{\cQtl}{\widetilde{Q}^c}

\newcommand{\eQtl}{\widetilde{Q}^e}

\newcommand{\rme}{{\rm e}}
\newcommand{\nAcal}{{\cal A}_0}
\newcommand{\Acal}{{\cal A}}
\newcommand{\tilJ}{ \tilde{\mathcal{J}} }
\def\anp#1#2#3{Annals Phys. {\bf #1}, #2 (#3)}
\def\arnps#1#2#3{Ann.\ Rev.\ Nucl.\ Part.\ Sci.\  {\bf #1}, #2 (#3)}
\def\atmp#1#2#3{Adv. Theor. Math. Phys. {\bf #1}, #2 (#3)}

\def\cqg#1#2#3{Class.\ Quant.\ Grav.\ {\bf #1}, #2 (#3)}
\def\epjc#1#2#3{Eur.\ Phys.\ Jour.\ C{\bf #1}, #2 (#3)}
\def\ibid#1#2#3{{\it ibid.} {\bf #1}, #2 (#3)}

\def\jhep#1#2#3{Jour. High Energy Phys. {\bf #1}, #2 (#3)}

\def\npa#1#2#3{Nucl. Phys. A {\bf #1}, #2 (#3)}
\def\npb#1#2#3{Nucl. Phys. B {\bf #1}, #2 (#3)}

\def\plb#1#2#3{Phys. Lett. B {\bf #1}, #2 (#3)}
\def\prc#1#2#3{Phys. Rev. C {\bf #1}, #2 (#3)}
\def\prd#1#2#3{Phys. Rev. D {\bf #1}, #2 (#3)}
\def\prl#1#2#3{Phys. Rev. Lett. {\bf #1}, #2 (#3)}
\def\phr#1#2#3{Phys. Rep. {\bf #1}, #2 (#3)}

\def\rpp#1#2#3{Rept. Prog. Phys. {\bf #1}, #2 (#3)}
\def\rmp#1#2#3{Rev. Mod. Phys. {\bf #1}, #2 (#3)}

\begin{document}
\preprint{KUNS-2216}
\pacs{11.10.Wx, 12.38.Mh, 25.75.-q}
\title{Hard thermal loops, to quadratic order,
in the background of a spatial 't Hooft loop}
\author{Yoshimasa Hidaka$^{a}$
and Robert D. Pisarski$^{b}$}
\affiliation{
$^a$Department of Physics, Kyoto University, Sakyo-ku, Kyoto 606-8502, Japan\\
$^b$Department of Physics, Brookhaven National Laboratory, Upton, NY 11973, USA\\
}
\begin{abstract}
We compute the simplest hard thermal loops 
for a spatial 't Hooft loop in the deconfined phase of
a $\SUN$ gauge theory.
We expand to quadratic order about a constant background field
$A_0 = Q/g$, where $Q$ is a diagonal, color matrix
and $g$ is the gauge coupling constant.
We analyze the problem in sufficient generality that the techniques
developed can be applied to compute transport properties in a
``semi''-Quark Gluon Plasma.
Notably, computations are done using the double line notation at finite $\Nc$.
The quark self-energy is a $Q$-dependent thermal mass squared,
$\sim g^2 \, T^2$, where $T$ is the temperature,
times the same hard thermal loop
as at $Q=0$.  The gluon self-energy involves two pieces:
a $Q$-dependent Debye mass squared, 
$\sim g^2 \, T^2$, times the same hard thermal
loop as for $Q=0$, plus a new hard thermal loop,
$\sim g^2 \, T^3$, 
due to the color electric field generated by a spatial
't Hooft loop. 
\end{abstract}
\date{\today}
\maketitle

\section{Introduction}
\label{introduction}

The spectacular success of the experiments at the Relativistic Heavy
Ion Collider (RHIC) at Brookhaven has invigorated the
study of gauge theories at a nonzero temperature, $T$
\cite{whitepaper,strong}.  Experiment
clearly shows that an understanding of quantities in thermal equilibrium
are not sufficient to understand the data, and that one also needs
quantities near thermal equilibrium, especially transport coefficients.

In this paper we compute
for what appears to be an unrelated problem: the real time response
functions for a spatial 't Hooft loop \cite{thooft1,thooft2}.  
A Wilson loop, 
${\rm tr} \, {\cal P} \exp( \, i g \oint_{\cal C} A_\mu \,dx^\mu)$,
represents the propagation of
a test electric charge along the path $\cal C$, and measures
the response to magnetic flux.  Similarly, a 't Hooft loop introduces
a test magnetic charge along a given path, and measures the response 
to electric flux.  Their behavior is dual to one another.
At zero temperature in a $\SUN$ gauge theory without quarks,
the condensation of magnetic charges confines electric charge, so
the Wilson loop has area behavior, and the 't Hooft loop, perimeter.
Conversely, at temperatures above that for deconfinement,
magnetic charges do not condense, and electric charge is not confined;
hence a (thermal) Wilson loop 
has perimeter behavior, and the (spatial) 't Hooft loop, area \cite{altes1}.

The 't Hooft loop does not have a simple representation in terms of the
vector potential for the gauge field, but 
in the deconfined phase, the area behavior of the spatial 't Hooft loop
can be simply understood \cite{altes1} as a $\ZN$ interface
\cite{interface1,altes2,altes3,linear_ZN,interface_lattice,real_time_lattice}.
Without dynamical quarks, a $\SUN$ gauge theory 
has $\Nc$ equivalent vacua, which differ by global $\ZN$ transformations
from one another \cite{thooft1,thooft2}.  To probe this, 
take a box which is long in one spatial direction, say of
length $L$ in $z$, and let the two ends of the box differ by
a $\ZN$ transformation.  Thus at one end of the box, $z = 0$,
the Wilson line in the imaginary time direction,
${\cal P} \exp(\, i g \int A_0 \, d\tau)$ equals the unit
matrix; at the other end of the box, $z= +L$, the Wilson line
is a $\ZN$ phase, $\exp(2 \pi i/N)$, times the unit matrix.
These boundary conditions can be imposed by introducing a background field
for the timelike component of the vector potential 
\cite{interface1,altes2,altes3,linear_ZN,interface_lattice,real_time_lattice,smilga,earlyA,earlyB,imag_chem_pot,background}:
\begin{equation}
A_0(z) = \frac{1}{g}  \; Q(z) \;  ,
\label{ansatz}
\end{equation}
where $Q$ is a diagonal matrix in color space, 
and $g$ is the gauge
coupling constant.  The matrix $Q(z)$ is then chosen to vary
so that a $\ZN$ interface, centered at $z=+L/2$, forms.
Only the ends of the box represent allowable vacua, so
a nonzero color electric field is generated along the $z$-direction,
$E_z \sim \partial_z Q(z)$,
and the configuration has nonzero action.  By construction,
the interface is independent of the $x$ and $y$ directions, and so
the action is proportional to the transverse area.  This $\ZN$ interface
is then equivalent to a spatial 't Hooft loop, in the plane of $x$ and
$y$, at $z = +L/2$ \cite{altes1}.

The action for the $\ZN$ interface can be computed in weak coupling,
and reduces to a tunneling problem in one dimension \cite{interface1}.
There is no potential for $Q$
classically, but one is generated at one-loop order,
and so the action for the associated instanton is not
$\sim 1/g^2$, but $\sim 1/\sqrt{g^2}$ \cite{interface1,resurgent}.
A derivative expansion can be used to compute, because along the
$z$-direction, the width of the interface is proportional
to the inverse Debye mass, $\sim 1/(g T)$, which is large relative
to the typical thermal correlation length for massless fields, 
$\sim 1/(2 \pi T)$.  Consequently, at the outset one computes
for a field $Q(z)$ which is constant in $z$, as
effects from 
the variation in $z$ enter through corrections which are of higher
order in $g$.  Corrections to the interface tension 
have been carried out to 
$\sim g^3$ \cite{altes2}, and are underway to $\sim g^4$ \cite{altes3}.
This is to be compared with the free energy (where $Q = 0$), which has
been computed to $\sim g^6$ 
\cite{braaten,perturbative,thermal_review,resum,pert_coupling}.

The interface tension is measurable through numerical simulations
on the lattice \cite{interface_lattice}.  This includes simulations
which model the behavior in real time \cite{real_time_lattice}.
In this paper we address a related problem semiclassically, by
computing the simplest real time response functions, for a spatial
't Hooft loop, in weak coupling.  To do this,
we expand to quadratic order
about the background field in Eq.~(\ref{ansatz}), taking $Q$ to be a constant
matrix, 
and analytically continue to real time.   As for quantities computed
in equilibrium, near equilibrium there is a natural division between
momenta which are ``soft'', with components of order the Debye
mass, $\sim g T$, and ``hard'', $\sim T$.  The simplest real time
response functions are the quark and gluon self-energies, computed
in the hard thermal loop approximation, for soft external momenta
\cite{htlA,htlB,lebellac}.  

We perform this computation in order to develop techniques which will enable
us to address a problem of much broader interest.
Resummations of perturbation theory
appear to break down by temperatures several times the critical
temperature \cite{thermal_review,resum}.  The obvious guess is to
assume that since $\Tc \sim 150-200$~MeV in a gauge theory, that the
theory has entered a nonperturbative regime by this point, where the
QCD coupling is large \cite{strong}.  
While the former must be true --- confinement cannot be
seen in perturbation theory --- computations
of an effective theory for the pressure find that
the coupling is {\it moderate} even at $\Tc$, with $\alpha_s^{\rm eff}(\Tc)
= g^2_{\rm eff}(\Tc)/(4 \pi) \sim 0.3$ \cite{pert_coupling}.  This is because 
in imaginary time, the typical
``energies'' are large, multiples of $2 \pi T$ \cite{braaten}.

Why, then, does deconfinement occur at moderate coupling?  
Deconfinement is an ordering of global $\ZN$ spins, and is measured
by the trace of the thermal Wilson line, which is the Polyakov loop
\cite{thooft1,thooft2}.  
In the fundamental representation, without quarks
the expectation value of the Polyakov loop
vanishes below $\Tc$, and approaches
one at high temperature.  Since the Polyakov loop is
not equal to one whenever $Q \neq 0$,
one way to model the region where the Polyakov loop is not near one
is to assume a nontrivial distribution of $Q$'s
\cite{loop1,loop2,loop3,loop4,lattice_effective,pnjl,real_time_lattice},
which in Ref. \cite{hidaka} we term
a ``semi'' Quark-Gluon Plasma (semi-QGP).
At least at infinite $\Nc$, it is easy to model the confined phase,
as a distribution which is flat in $Q$.  This implies that the
expectation value of the Polyakov loop in any nontrivial representation
vanishes, whether or not the loop carries $\ZN$ charge.
This is also consistent with how the 't Hooft loop must change 
near $\Tc$.
At high temperature, $\ZN$ interfaces are rare, and in infinite
volume the theory lies in one $\ZN$ domain.  As 
$T \rightarrow \Tc^+$, though,
the interface tension decreases, $\ZN$ domains become
plentiful, and the $\ZN$ spins are disordered.
The decrease of the $\ZN$ interface tension near $\Tc$ has been confirmed 
on the lattice \cite{interface_lattice}.

A semi-QGP can be shown to occur in one unphysical
limit.  Let the spatial volume be a sphere of hadronic dimensions,
so small that by asymptotic freedom, the coupling constant runs
to a very small value \cite{small_sphere1}.
Although systems at finite volume cannot have phase transitions,
they can if the number of colors is infinite.
If $R$ is the radius of the sphere,
then even when $g^2 = 0$, at infinite $\Nc$
there is a deconfining phase transition, of first order, when
$\Tc = c/R$, where $c = 1/\log(2 + \sqrt{3})$ \cite{small_sphere1}.
At $\Tc^+$, the expectation value of the 
(renormalized) Polyakov loop, in the fundamental
representation, is exactly $1/2$.
The $Q$-distribution for the constant mode on the sphere reduces to a type of
matrix model, which can be computed analytically 
about $\Tc$ \cite{small_sphere1}.  Further, since all resonances are
of zero width at infinite $\Nc$, 
the Hagedorn temperature is a precise quantity.
On a small sphere at $\Nc = \infty$, at zero coupling
the Hagedorn temperature coincides with $\Tc$.
Perturbative corrections move $\Tc$ below the Hagedorn temperature,
and the loop at $\Tc^+$ away from $1/2$, by an amount 
$ \sim (g^2(R) \, \Nc)^2$ \cite{small_sphere2}.
Presumably, the expectation value of
the (renormalized) Polyakov loop, in the fundamental representation,
goes from $1/2$ at $\Tc^+$, to near one, by temperatures which
are a few times $\Tc$. This is then the semi-QGP on a small sphere.

Of course this might be an artifact of working
on a small sphere at infinite $\Nc$.
In a large spatial volume, one must look to numerical simulations
on a lattice.  In any volume, the Wilson line, and so the Polyakov loop
has ultraviolet divergences, so that the bare loop vanishes in the
continuum limit.  A nonzero value in the continuum limit is obtained
after a type of mass renormalization
\cite{loop2,ren_loop1,ren_loop2,renloop_lattice1,renloop_lattice2,renloop_lattice3,ghk}.
On a small sphere, when $g^2(R) \ll 1$,
one can renormalize the loop perturbatively.  In a large volume,
on the lattice there are two methods of renormalizing the
Polyakov loop \cite{loop2,ren_loop1}, which now agree up to the numerical
accuracy \cite{ghk}.
The most precise measurements are for
a ${\rm SU}(3)$ gauge theory without quarks \cite{ghk}.  
From Fig.~1 of \cite{ghk}, in the triplet representation the 
expectation of the renormalized
Polyakov loop vanishes below $\Tc$,
and is $\sim 0.5$ at $\Tc$.  It then rises to 
it rises rapidly, and is $\sim 0.9$ by $2.0 \, \Tc$.
It then rises slowly, reaching $\sim 1.1$ by $\sim 4 \, \Tc$.
From $4$ to $12 \, \Tc$, its expectation value is flat.
This suggests that there is a nontrivial $Q$-distribution about
$\Tc$, which is relevant up to temperatures which are $2 - 4 \, \Tc$.
Above $4 \, \Tc$, any nonperturbative effects from the $Q$-distribution
appear negligible, consistent with the success of
resummations of perturbation
theory \cite{resum,pert_coupling}.  In a ${\rm SU}(3)$ gauge theory
with quarks, the simulations are of more limited accuracy,
but a similar picture emerges \cite{renloop_lattice3}.  
The principal difference
is that the expectation value of the renormalized triplet loop is
nonzero even below $\Tc$ \cite{renloop_lattice3}.

At present, it is not known what the $Q$-distribution is in a gauge
theory, even without quarks.  This would correlate the pressure with
the expectation value of the (renormalized) Polyakov loop(s).  Such
a distribution might be obtained from numerical simulations, in both
the original and an effective theory \cite{lattice_effective}.

Even without knowing the full $Q$-distribution, though, one can take
the first steps toward the response functions in real time, which we
do here.  For either a $Q$-distribution, or a 't Hooft loop, one begins
by computing the quadratic fluctuations about the background field
in Eq.~(\ref{ansatz}).  We then analytically continue the fluctuations,
computed in imaginary time, to real time \cite{lebellac,furuuchi}.
(For a numerical approach to quantities in 
real time, see \cite{real_time_lattice}.)

We find that the quark self-energy is a trivial extension of that
for $Q=0$: it is equal to a thermal mass squared, $\sim g^2 \, T^2$, times
the same hard thermal loop as in zero field.  The gluon self-energy
is different, though.  Besides the usual hard thermal loop,
proportional to a thermal mass squared, $\sim g^2 \, T^2$, there is
a new piece, $\sim g^2 \, T^3$.   The function is similar to other hard 
thermal loops, but is novel.  We suggest that it arises because of the
background color electric field in a 't Hooft loop.  

The paper is organized as follows:  In Sec.~\ref{sec_basis}, we 
discuss how the double line notation,
which is standard at large $\Nc$ \cite{thooft_planar}, 
can also be used easily at finite $\Nc$ \cite{cvitanovic}.  
Our discussion is elementary, but is absolutely essential to being
able to compute in an arbitrary background field with $A_0 \neq 0$,
Eq.~(\ref{ansatz}).

Section~\ref{sec_background_field}
introduces the background field calculation for constant $A_0$.
The perturbative rules in the 't Hooft basis 
are given in Sec.~\ref{subsec_background}.
These are nothing more involved than the usual
perturbative rules, with a simple ``shift'' in the energies, $\sim Q$.
Sec.~\ref{sec_mixed_props} gives expressions, useful for 
calculation, in terms of a ``mixed'' representation, 
working with spatial momenta and Euclidean time.
In Sec.~\ref{sec_analytic_cont}, we 
follow Furuuchi \cite{furuuchi} and discuss how to obtain scattering
amplitudes, in real time, from those computed in imaginary time.

The computation of the hard thermal loop
in the quark self-energy is given in Sec.~\ref{sec_quark_self_energy}.  
We go through this example in some detail to develop familiarity
with computing diagrams when $Q \neq 0$.
The hard thermal loop in the gluon self-energy is computed in
Sec.~\ref{sec_gluon_self_energy}.  In Sec.~\ref{subsec_gluon_htls_tadpoles}
we discuss the $Q$-dependence of the hard thermal loops
in two tadpole diagrams, which are
independent of the external momenta.  
In Sec.~\ref{subsec_gluon_htls_landau} we consider the hard thermal
loops which arise from diagrams which exhibit with Landau damping.
With these examples in hand, computing the
one point gluon function, in Sec.~\ref{sec_one_point_gluon}, the quark contribution to the
gluon self-energy,
Sec.~\ref{quarkcontgluonselfenergy}, 
and that of ghosts and gluons to the gluon
self-energy, Sec.~\ref{gluoncontgluonselfenergy},
is relatively straightforward.

In an appendix we draw some distinctions on the differences between
$\ZN$ and $\U1$ interfaces \cite{loop4}.  

\section{The double line notation at finite $\Nc$.}
\label{sec_basis}

In order to compute efficiently, it is useful to have a convenient
basis for the generators of $\SUN$.  In this section we 
follow Cvitanov\'ic  \cite{cvitanovic} to show how
the usual double line notation, which is familiar at
large $\Nc$ \cite{thooft_planar}, is also natural
at small $\Nc$.

Our purpose here is to establish the notation that 
we need to compute in the presence of a constant, 
background field for $A_0$, Eq.~(\ref{ansatz}).  
We note that at $\Nc = \infty$, Aharony {\it et al.} computed 
the free energy with $A_0 \neq 0$ to three-loop order,
$\sim g^4$, on a small sphere \cite{small_sphere1}.  
At finite $\Nc$, recently Korthals-Altes used the 
double line notation to compute
the free energy for $A_0 \neq 0$ to $\sim g^2$ in supersymmetric theories \cite{altes3}.

The standard choice for the generators of a gauge group, $\lambda^A$, is 
to take a complete and orthonormal basis,
\beq
\trace  \left( \lambda^A \; \lambda^B 
\right)= \frac{1}{2} \; \delta^{A B} \; ;
\label{cartan_normalization}
\eeq
$A$ and $B$ refer to adjoint indices, which for $\SUN$ run from
$A,B=1,2,\ldots,(\Nc^2 -1)$.  

We denote indices in the fundamental representation by 
$a,b\ldots= 1,2,\ldots, \Nc$.  
Indices in the adjoint representation are then denoted
by a pair of fundamental indices, $(ab)$.  
The basic quantity which we need is a projection operator which ties 
together upper and lower adjoint indices:
\beq
\proj^{a b}_{c d} 
= \delta^a_{c} \; \delta^b_{d} - \frac{1}{\Nc}\;
\delta^{a b} \; \delta_{c d} \; .
\label{projector}
\eeq
For an arbitrary matrix $M$,
adjoint indices are raised and lowered by
flipping the order in the pair:
$ M^{a b} = M_{b a}$.
The indices flip because off-diagonal generators are ladder operators.
With this convention, the projection operators between a pair
of upper indices, or a pair of lower indices, are 
\beq
\proj^{a b}_{c d} = 
\proj^{a b, d c} = 
\proj_{b a , c d} \; .
\eeq
It is obvious that Eq.~(\ref{projector}) represents a projection operator
\beq
\proj^{a b}_{e f} \;\; \proj^{e f}_{c d} =
\proj^{a b}_{c d} \; .
\label{product_proj}
\eeq
While we have to be careful in lowering and raising adjoint indices,
we can raise or lower single indices without concern,
$\delta^{a b} = \delta^a_b = \delta_{a b}$.
The second term in Eq.~(\ref{projector}) ensures that it is traceless
in either of the two pairs of adjoint indices,
\beq
\delta_{a b} \; \proj^{a b}_{c d} = 
\delta^{c d} \; \proj^{a b}_{c d} = 0 \;.
\label{trace_condition}
\eeq

While we call $(a b)$ an adjoint index, 
this terminology is somewhat misleading.  $\SUN$ has
$\Nc^2 -1 $ independent generators, but there are obviously
$\Nc^2$ values for the index $(a b)$.  By using 
projection operators
(or in the terminology of \cite{cvitanovic}, invariant tensors), the resulting
basis is overcomplete, with one extra generator.

While we give explicit expressions for all quantities,
it is handy to use a diagramatric notation
\cite{thooft_planar,cvitanovic}.
For $\SUN$, lines always carry an arrow, with
fields in the fundamental representation represented by
a single line, and those in the adjoint, by a double line.  For
an adjoint index $(a b)$, we adopt the notation that 
for upper indices, $a$ is outgoing and $b$ ingoing; 
for lower indices, $a$ is ingoing and $b$ outgoing.
This reversal is necessary so that upper and lower indices
are contracted accordingly.

\begin{figure}
\includegraphics[width=0.35\textwidth]{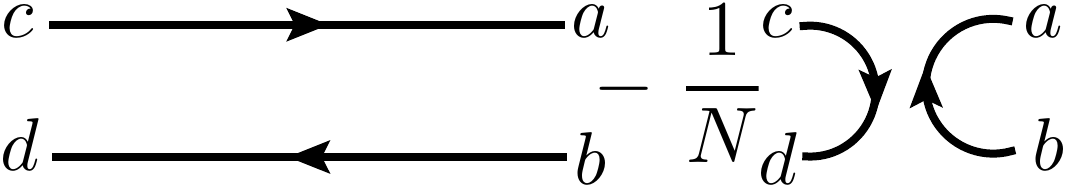}
\caption{Projection operator, $\mathcal{P}^{ab,dc}$.}
\label{fig_projector}
\end{figure}

Generally, any confusion with indices is dispelled by drawing the
corresponding diagram.
For example, the projection operator of Eq.~(\ref{projector})
is illustrated in Fig.~\ref{fig_projector},
and is drawn like a gluon propagator.

\begin{figure}
\includegraphics[width=0.5\textwidth]{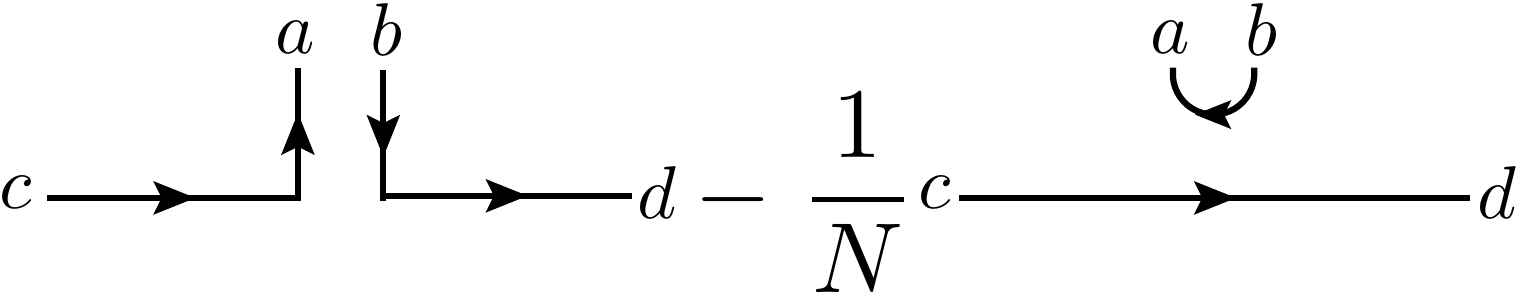}
\caption{Generator for $\SUN$, times $\sqrt{2}$.}
\label{fig_generator}
\end{figure}

With these conventions,
the generators of the fundamental representation are just projection 
operators,
\beq
(t^{a b})_{c d} = \frac{1}{\sqrt{2}} \; \proj^{a b}_{c d} \; .
\label{generator_projector}
\eeq
Note that the upper pair, $(a b)$, refers to the index for the adjoint
representation, while the lower pair, $(c d)$, refers to the components
of this matrix in the fundamental representation.  
This is illustrated in
Fig.~\ref{fig_generator}.  While this is also a projection operator,
we draw it differently from Fig.~\ref{fig_projector}, 
distinguishing between the adjoint indices, on top of the
diagram, and the matrix
indices for the fundamental representation, on the two sides.

As each generator is a projection 
operator, the trace of two generators is again
a projection operator:
\beq
\trace \; \left( t^{a b} \; t^{c d} \right)
= \frac{1}{2} \; \proj^{a b}_{e f}
\; \proj^{c d}_{f e} = \frac{1}{2} \; \proj^{a b,c d} \; .
\label{trace_gen}
\eeq
We now make an extended comment about the normalization of generators
which is implied by Eq.~(\ref{trace_gen}).  While mathematically
elementary, at least we found it confusing at first.

The off-diagonal
generators are the customary ladder operators 
of the Cartan basis.  That is, for $a \neq b$, 
they are normalized as in Eq.~(\ref{cartan_normalization}), i.e.,
\beq
\trace \left( t^{a b}\; t^{b a} \right) = \frac{1}{2} \; ;
\label{normalization}
\eeq
here $a$ and $b$ are fixed indices, with no summation convention.
In $\SUN$, there are $\Nc (\Nc -1)$ off-diagonal generators.

The only difference lies in the choice of the diagonal generators
(which is the Cartan subalgebra, the space of mutually commuting 
generators).  The Cartan basis includes one generator proportional to
\beq
t^{\Nc \Nc} = \frac{-1}{\Nc \sqrt{2}} \; 
\left(
\begin{array}{cc}
{\bf 1}_{\Nc-1} & 0      \\
0             & -(\Nc-1) \\
\end{array}
\right) \; ,
\label{highest_nc}
\eeq
where ${\bf 1}_{\Nc-1}$ is the unit matrix in $\Nc - 1$ dimensions.
In the Cartan basis, the corresponding matrix is
$\lambda^\Nc = -\sqrt{\Nc/(\Nc - 1)} \; t^{\Nc \Nc}$, where
the overall constant is required so that $\lambda^\Nc$ obeys
Eq.~(\ref{cartan_normalization}).
In the Cartan basis, the other diagonal
generators are like $t^{\Nc \Nc}$, but for
smaller $\Nc$.  For example, 
\beq
\lambda^{\Nc-1}
= \frac{1}{\sqrt{2 (\Nc - 1)(\Nc - 2)}} 
\; \left(
\begin{array}{ccc}
{\bf 1}_{\Nc-2} & 0      &0 \\
0             & -(\Nc-2) & 0\\
0             & 0      & 0 \\
\end{array}
\right) \; ,
\label{second_cartan_matrix}
\eeq
and so on.
We denote the $\Nc - 1$ diagonal generators in the Cartan basis as
$\lambda^a$, $a = 2,3,\ldots, \Nc$.
While orthonormal, this basis
clearly treats the different diagonal elements on an unequal footing,
with the $\Nc^{\rm th}$ element occupying a privileged position.

In contrast, for the double line basis
the diagonal generators are just
permutations of one another:
start with Eq.~(\ref{highest_nc}), and simply
shuffle where the factor of $-(\Nc-1)$ lies along the diagonal,
e.g., 
\beq
t^{1 1} = \frac{-1}{\Nc \sqrt{2} } \; 
\left(
\begin{array}{cc}
-(\Nc-1) & 0      \\
0             & {\bf 1}_{\Nc-1} \\
\end{array}
\right) \; ,
\label{lowest_nc}
\eeq
and so on.  The $t^{11}, \ldots, t^{\Nc \Nc}$ are a set of $\Nc$
diagonal generators, which manifestly do not treat any diagonal element
different from any other.
This is only possible only because they are not independent, that is, 
their sum vanishes,
\beq
\sum_{a = 1}^{\Nc} \; t^{a a} = 0 \; .
\label{sum}
\eeq

Consider the example of two colors, where the double line basis has four 
generators.  There are two off-diagonal, ladder generators,
\beq
t^{1 2} = \frac{1}{\sqrt{2}} \; \left(
\begin{array}{cc}
0 & 1 \\
0   & 0 \\
\end{array}
\right) \;\; , \;\;
t^{21} = \frac{1}{\sqrt{2}} \; \left(
\begin{array}{cc}
0 & 0 \\
1   & 0 \\
\end{array}
\right) \; \; . \;\;
\label{thooft_two_color_basisA}
\eeq
For the diagonal generators, from Eqs.~(\ref{projector}) and (\ref{generator_projector}) there are two contributions,
\beq
t^{11} = 
\frac{1}{\sqrt{2}} 
\left(
\; \left( \begin{array}{cc}
1 & 0 \\
0   & 0 \\
\end{array} \right) \; - 
\frac{1}{2} \; \left( \begin{array}{cc}
1 & 0 \\
0   & 1 \\
\end{array} \right) 
\right) =
\frac{1}{2 \sqrt{2}} \; \left(
\begin{array}{cc}
1 & 0 \\
0   & -1 \\
\end{array}
\right) = - t^{22}\; .
\label{thooft_two_color_basisB}
\eeq
As expected, $t^{1 1}$ is proportional to the Pauli matrix $\sigma^3$,
but the constant appears wrong.  While 
the ladder generators
are normalized as in Eq.~(\ref{normalization}),
${\rm tr}\left( t^{12} t^{21} \right) = 1/2$, the diagonal generators
satisfy
${\rm tr} \left(t^{11} \right)^2 = {\rm tr} \left(t^{22}\right)^2  = 1/4$, 
instead of $1/2$;
further, 
${\rm tr} \left( t^{11} t^{2 2} \right) = - 1/4 $.

This normalization is correct, and 
arises because the basis is overcomplete.
From Eq.~(\ref{trace_gen}), the trace of a product of two generators
is itself a generator.  Thus the trace of a given diagonal generator,
squared, is
\beq
\trace \; (t^{aa})^2 = \frac{1}{2} \left( 1 -
\frac{1}{\Nc} \right) \; ,
\label{nonorth1}
\eeq
with no summation over $a$; for $\Nc = 2$, this $= 1/4$.
Further, the trace between two different diagonal generators is nonzero:
\beq
\trace \; t^{aa} \; t^{b b} = \frac{1}{2} \; \left( - \;
\frac{1}{\Nc}\right) \; ;
\label{nonorth2}
\eeq
with no summation over $a$ or $b$, and $a \neq b$; 
for $\Nc = 2$, this $=-1/4$.
Thus in this basis, the (peculiar) normalization of the diagonal
generators arises because the generators are projection operators.
Of course the diagonal and off-diagonal generators are orthogonal
to one another in the usual manner.

\begin{figure}
\includegraphics[width=0.6\textwidth]{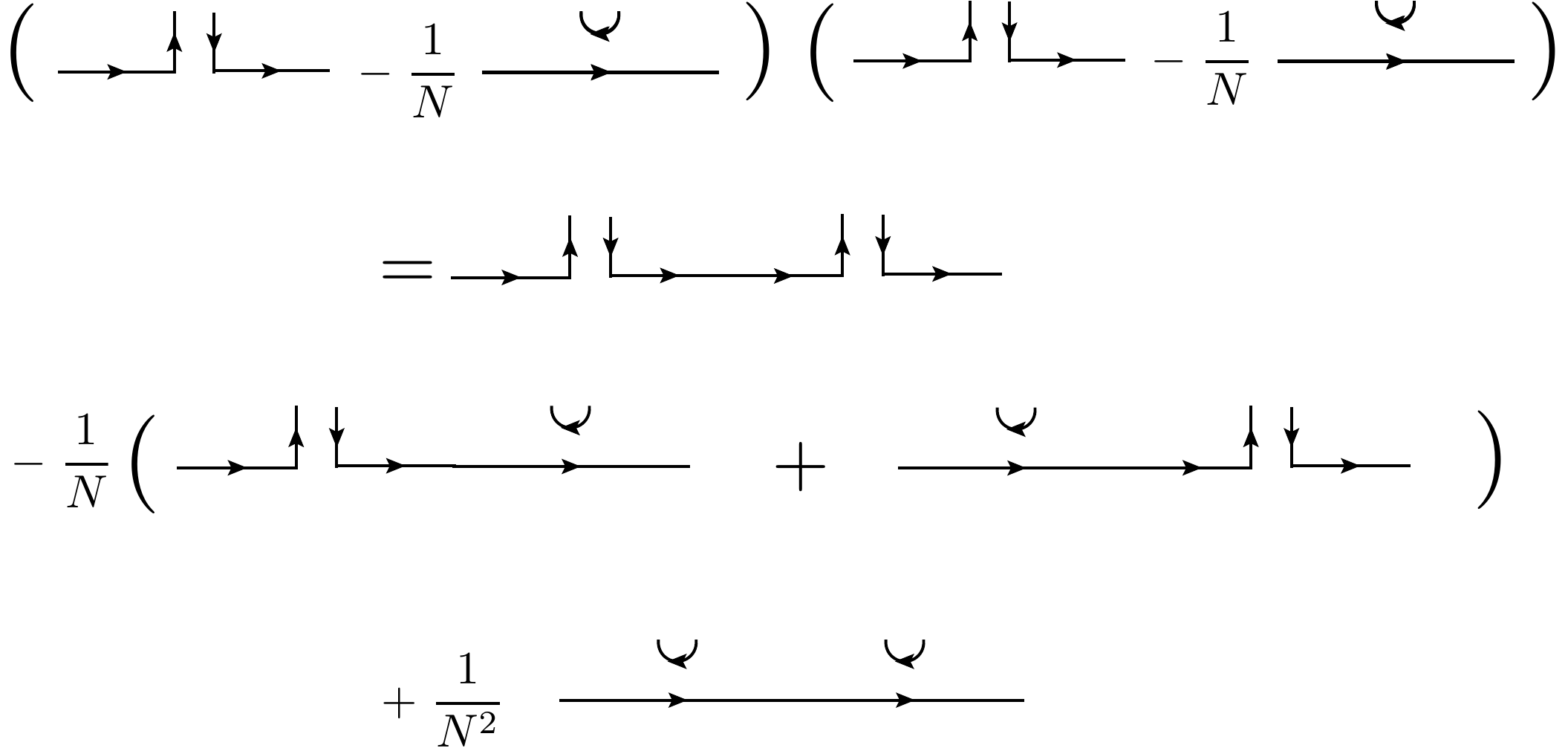}
\caption{The product of two generators, times two.}
\label{fig_product_two_generators}
\end{figure}

In the absence of a background field,
that the Cartan basis chooses a preferred direction amongst
the diagonal generators is of
no concern.  The Cartan basis is also convenient when computing the
properties of a $\ZN$ interface \cite{interface1},
since then the background field is along $t^{\Nc \Nc}$,
and treating the $\Nc^{\rm th}$ diagonal element as special is 
natural.

In the presence of an {\it arbitrary} background field, Eq.~(\ref{ansatz}),
though, where $Q$ is a diagonal matrix,
$(Q)_{ab} = Q^a \delta_{a b}$, Eq.~(\ref{background_field_ansatz}),
the double line basis is more useful.
In particular, all covariant derivatives are simple.
In the fundamental representation, $Q$ acts linearly upon fields $\psi$.
Then $\psi$ is like a column vector,
so if $\psi_a$ is the $a^{\rm th}$ element, 
\beq
Q \; \psi_a =  Q^a \; \psi_a
\; .
\label{fundamental_action_Q}
\eeq
In the adjoint representation, the covariant derivative involves a
commutator.  The commutator of
$Q$ with any generator, though, is just that generator times a difference
of $Q$'s:
\beq
[Q, t^{a b}] = (Q^a - Q^b) \; t^{a b} \equiv Q^{a b} \; t^{a b} \; .
\label{adjoint_action_Q}
\eeq
This is clear: if $a=b$, $t^{a b}$ is a diagonal matrix, and the commutator
of $t^{a a}$ with another diagonal matrix, $Q$, vanishes.  If $a \neq b$,
only the first term in $(t^{a b})_{c d}$, $\sim \delta^a_c \delta^b_d$,
Eqs.~(\ref{projector}) and (\ref{generator_projector}), 
contributes, to give Eq.~(\ref{adjoint_action_Q}).  We introduce
the notation $Q^{a b} = Q^a - Q^b$, which we shall use extensively.

As a consequence of Eqs.~(\ref{fundamental_action_Q}) and 
(\ref{adjoint_action_Q}), we find in Sec.~\ref{subsec_background}
that with the double line basis,
perturbation theory for $Q \neq 0$ is a
{\it trivial} generalization of that for $Q=0$: just a constant,
albeit color dependent, shift in the energies.
Energies which carry color indices were introduced when
the determinant in a background constant $A_0$ field was first computed, in
Appendix D of Ref.~\cite{earlyA}.

The double line basis is useful in other ways.  While admittedly perverse
for two colors, for three or more colors
it is a very efficient
means of deriving various identities amongst generators of the gauge group.
For example, the product of two generators is 
\beq
\left(t^{a b} \; t^{c d}\right)_{e f}
= \frac{1}{2} \; \proj^{a b}_{e g} \; \proj^{c d}_{g f} 
\label{product_two_gens}
= \frac{1}{2} \left( 
\delta^a_e \, \delta^{b c} \, \delta^d_f
- \frac{1}{\Nc} \left( \delta^a_e \, \delta^b_f \, \delta^{c d}
+ \delta^{a b} \, \delta^c_e \, \delta^d_f \right)
+ \frac{1}{\Nc^2} \, \delta^{a b} \, \delta^{c d} \, \delta_{e f}  
\right) \; ,
\eeq
as illustrated in Fig.~\ref{fig_product_two_generators}.
This certainly shows how writing down all of the indices is 
more tedious than just drawing the corresponding diagram.
By tying the sides of the diagram
together, representing summation over the matrix
indices, we obtain the normalization conditions above, 
Eqs.~(\ref{nonorth1}) and (\ref{nonorth2}).  By tying the adjoint indices
together on top of the diagram, we also obtain the familiar identity:
\beq
\sum_{a,b=1}^\Nc \; \left(t^{a b} \; t^{b a}\right)_{c d} = 
\frac{\Nc^2 -1}{2 \Nc} \; \delta_{c d} \; .
\label{sum_squared_gens}
\eeq
In principle, we really should contract the adjoint indices with
a projection operator.  Since the generators are traceless, though,
Eq.~(\ref{sum}), the projection operator reduces to ordinary Kronecker deltas.

The product of three arbitrary generators can be written out,
similar to Fig.~\ref{fig_product_two_generators}.  
By drawing diagrams, it is easy obtaining the standard relation,
\beq
\sum_{c, d =1}^\Nc \; t^{c d} \; t^{a b} \; t^{d c}
= - \frac{1}{2 \Nc} \; t^{a b} \; ,
\eeq
where we leave the matrix indices implicit.

\begin{figure}
\includegraphics[width=0.35\textwidth]{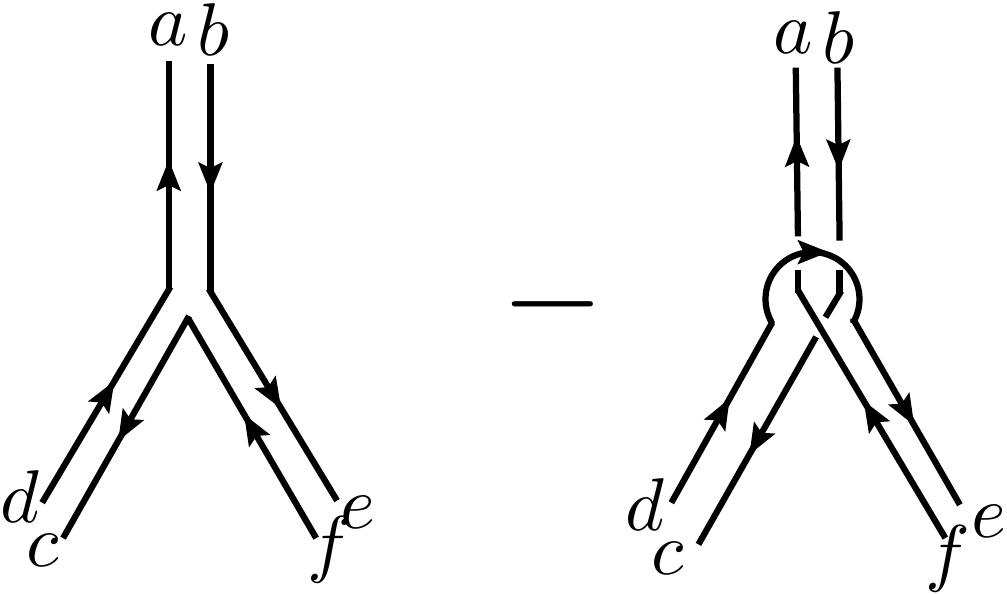}
\caption{Structure constant for $\SUN$, times $- i \sqrt{2}$.}
\label{fig_three_gluon_vertex}
\end{figure}

More useful is to take a trace of the product of three generators.
The antisymmetric combination is proportional to the structure
constant of the group,
\beq
[t^{a b}, t^{c d}] = i \sum_{e,f = 1}^\Nc \; f^{(ab, cd, ef)} \; t^{f e} \; ,
\label{structure_const}
\eeq
and is simple,
\beq
f^{(ab, cd, ef)} =  \frac{i}{\sqrt{2}}
\left( \delta^{a d} \, \delta^{c f} \, \delta^{e b} 
- \delta^{a f} \, \delta^{c b} \, \delta^{e d}  \right) \; ,
\label{three_gluon_vertex}
\eeq
as illustrated in Fig.~\ref{fig_three_gluon_vertex}.
In this basis the structure constants satisfy the identity:
\beq
\sum_{e,f,g,h = 1}^\Nc \; f^{(a b, e f, g h)} \; f^{(c d, f e, h g)}
\; = \; \Nc \; {\cal P}^{a b, c d} \; .
\label{sum_fs}
\eeq

\begin{figure}
\includegraphics[width=0.6\textwidth]{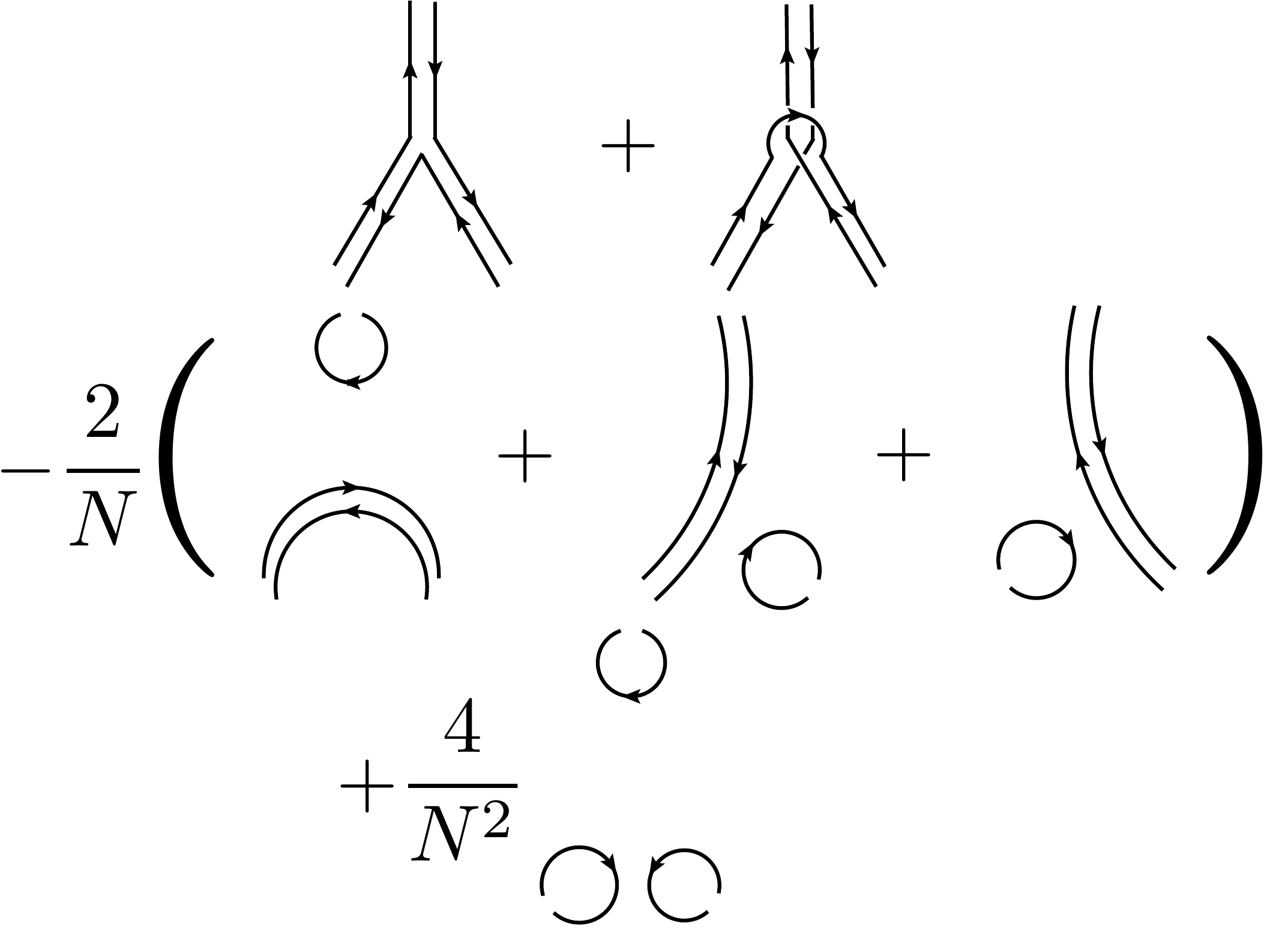}
\caption{Symmetric structure constant, times $\sqrt{2}$.}
\label{fig_symmetric_tensor}
\end{figure}

It is also easy to draw the diagram for the symmetric structure
constant, $d^{(a b, c d, ef)}$,
\beq
\begin{split}
d^{(a b, c d, ef)} = &2 \;\trace \left( t^{a b} \{t^{c d},t^{e f}\}\right)\\
= &
\frac{1}{\sqrt{2}} \;
\Bigl( \delta^{a d} \delta^{c f} \delta^{e b} 
+ \delta^{a f} \delta^{c b} \delta^{e d}  
-\frac{2}{\Nc} \; \left(
\delta^{a b} \delta^{c f} \delta^{e d}
+ \delta^{a d} \delta^{c b} \delta^{e f} 
+ \delta^{a f} \delta^{c d} \delta^{e b}
\right) \\
&\qquad+ \frac{4}{\Nc^2} \;
\delta^{a b} \delta^{c d} \delta^{e f} \Bigr) \; ,
\end{split}
\eeq
as illustrated in Fig.~\ref{fig_symmetric_tensor}.

For higher representations of the group, instead of diagrams with just
two lines, one obtains diagrams with many lines, or
``birdtracks''.  For a careful discussion
of the classification of arbitrary representations of Lie groups by
means of birdtrack diagrams, see Cvitanov\'ic \cite{cvitanovic}.

\section{Computing in background field gauge}
\label{sec_background_field}
\subsection{Propagators in a background $A_0$ field}
\label{subsec_background}

In this section we develop the perturbative rules in the appropriate background
field \cite{background}.  At tree level the Lagrangian is
\beq
{\cal L} = \frac{1}{2} \; \trace \left( G^2_{\mu \nu} \right) \; 
+ \overline{\psi}\left( \Slash{D} +m \right)\psi \; .
\label{classical_lag}
\eeq
We assume there are $N_f$ flavors of quarks, $\psi$, in the
fundamental representation of the gauge group;
the covariant derivative in that representation
is $D_\mu=\partial_\mu-igA_\mu$.
The field strength tensor
$G_{\mu\nu}= [D_\mu,D_\nu]/(-ig)=\partial_\mu 
A_\nu -\partial_\nu A_\mu -ig[A_\mu,A_\nu]$.  The covariant derivative
in the adjoint representation is $D_\mu = \partial_\mu-ig [A_\mu, \; . \;]$.
We work in Euclidean spacetime, with a positive metric.
The gamma matrix is Hermitian and satisfies $\{\gamma_{\mu},\gamma_{\nu}\}=2\delta_{\mu\nu}$.

We expand about a background field $\acl_\mu$,
\beq
A_\mu = \acl_\mu + B_\mu \; ,
\label{background_expand}
\eeq
where $B_\mu$ denotes the fluctuation.
The classical covariant derivative is then
$D^\text{cl}_\mu = \partial_\mu-ig \acl_\mu$, etc.
The gauge fixing and ghost terms are chosen to be
those for background field gauge, with gauge fixing parameter $\xi$:
\beq
{\cal L}_\text{gauge} = 
\frac{1}{\xi} \; \trace \left(D^\text{cl}_{\mu}
B_\mu\right)^2 \; -2 \;
\trace\left( \bar{\eta}  D^\text{cl}_\mu D_\mu  \eta \right)\; ,
\label{gauge_fixing}
\eeq
with $\eta$ the ghost field.

The inverse propagators follow directly.
That for the quark field is $\Slash{D}^\text{cl} + m$.
The inverse propagator
for the ghost is $- (D^\text{cl}_\mu)^2$, while that for gluon
fluctuations, $B_\mu$, is
\beq
\begin{split}
 (\varDelta^\text{cl}_{\mu\nu})^{-1} 
& = \; - \left(D^\text{cl}_\lambda\right)^2 \, \delta_{\mu\nu}             
\; + \; D^\text{cl}_\nu \, D^\text{cl}_\mu \; - \; \frac{1}{\xi} \;
D^\text{cl}_\mu \, D^\text{cl}_\nu               
\; + \; i g \, [G^\text{cl}_{\mu\nu},\; . \;]   \\
& = \; - \left(D^\text{cl}_\lambda\right)^2 \,
\delta_{\mu \nu} \; + \; \left( 1 - \frac{1}{\xi} \right)
D^\text{cl}_\mu \, D^\text{cl}_\nu \; + \; 2 i g \, [G^\text{cl}_{\mu\nu},\; . \;] \; .
\end{split}
\label{field_strength}
\eeq

Most of our calculations are done assuming a background field which is
constant in spacetime.  Notice, however, that the 
last term in the inverse gluon propagator
is proportional to the field strength tensor of the background field.  
This will be important in understanding novel terms for gluon
hard thermal loops in the presence of an interface.

The covariant derivative in the fundamental representation
enters into the quark inverse propagator, while that in the adjoint
representation enters into the ghost and gluon inverse propagators.
We now compute at a nonzero temperature $T$ in the imaginary time
formalism, where the Euclidean time $\tau: 0 \rightarrow 1/T$.
The energies are then 
\begin{equation}
p_0 = 2 n \pi T , \;\; {\rm bosons} \;\;\;\; ; \;\;\;\;
\widetilde{p}_0 = (2 n + 1) \pi T , \;\; {\rm fermions} \; .
\end{equation}
We use a tilde for the energies and momenta of fermions, to distinguish
them from bosons.

We take the background field
as a constant, diagonal matrix for the timelike component of the vector
potential:
\beq
\acl_0 \; = \; \frac{1}{g} \; Q \;\;\; ; \;\;\;
(Q)_{a b} = Q^a \, \delta_{a b} \; ;
\label{background_field_ansatz}
\eeq
as an $\SUN$ matrix, the sum of the $Q^a$'s vanishes,
\beq
\sum_{a = 1}^{\Nc} Q^a = 0  \; .
\label{Qa}
\eeq

As discussed in Sec.~\ref{sec_basis}, the great virtue of the
double line notation is that the covariant derivatives in a field like
Eq.~(\ref{background_field_ansatz}) are trivial.
For fields in the fundamental representation,
the covariant derivative in the background field is
$
D_\mu^\text{cl}  \, \psi_a = - i \widetilde{P}_\mu^a  \, \psi_a \; ,
$
Eq.~(\ref{fundamental_action_Q}).
If this covariant derivative acts upon a quark field, 
$\widetilde{P}_\mu^a$ is a momenta with one color index,
\beq
\widetilde{P}_\mu^a = \left(\widetilde{p}_0 + Q^a,\vec{p}\; \right) \; .
\label{quark_momenta}
\eeq
Since the background field shifts the Euclidean energies, 
it is convenient to write
\beq
\aPtl_\mu = P_\mu + \aQtl \; , \;
\aQtl = Q^a + \pi T\; .
\label{shifted_quark_momenta}
\eeq
That is, we treat all momenta as bosonic, which we can 
easily do by just putting the change in the boundary condition
for fermions, versus bosons, into part of the background field.

The covariant derivative acts upon fields in the adjoint representation as
$
D_\mu^\text{cl}  \, t^{a b} = - i P_\mu^{a b}  \, t^{a b} \; ,
$ 
Eq.~(\ref{adjoint_action_Q}).  For bosonic fields,  
\beq
P_\mu^{a b} = (p_0 + Q^a - Q^b,\vec{p}\; ) 
= (p_0^{a b},\vec{p}) \; ,
\label{gluon_momenta}
\eeq
and involves an adjoint color index, $(a b)$.

\begin{figure}
\begin{align}
&\parbox{5.5cm}{\includegraphics[width=0.23\textwidth]{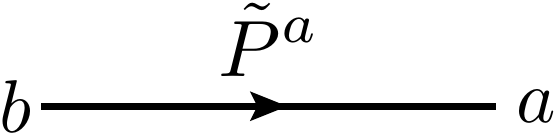}}
=  \; \frac{\delta^{ a b}}{ -i \!\! \not \!\!\,\aPtl +m} \; , \notag \\
&\parbox{8cm}{\includegraphics[width=0.4\textwidth]{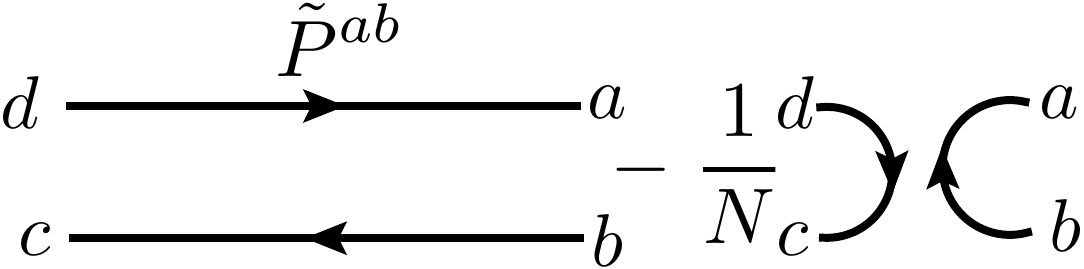}}
=  \; \frac{1}{(P^{a b})^2 } \; \proj^{a b, c d}  \; , \notag \\
&\parbox{8cm}{\includegraphics[width=0.4\textwidth]{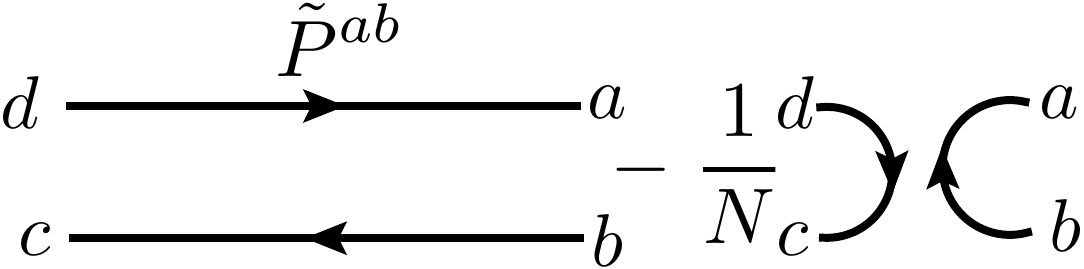}}
= \left( \delta_{\mu\nu} - (1 - \xi) \;
\frac{P^{a b}_\mu P^{a b}_\nu}{(P^{a b})^2} \right)
\; \frac{1}{(P^{a b})^2} \; \proj^{a b, c d} \; . \notag
\end{align}
\caption{Quark, ghost, and gluon propagators.}
\label{fig_propagators}
\end{figure}

To be explicit, the quark propagator is
\beq
\langle \; \psi^a(P) \; \overline{\psi}^b(-P) \;  \rangle
=  \; \frac{\delta^{ a b}}{ -i \!\! \not \!\!\,\aPtl +m} \; ;
\label{quark_prop}
\eeq
the ghost propagator,
\beq
\langle \;  \eta^{a b}(P) \; \overline{\eta}^{c d}(-P) \; \rangle
=  \; \frac{1}{(P^{a b})^2 } \; \proj^{a b, c d}  \; ;
\label{ghost_prop}
\eeq
and the gluon propagator,
\beq
\langle \; B_\mu^{a b}(P) \; B^{c d }_\nu(-P) \; \rangle
= \left( \delta_{\mu\nu} - (1 - \xi) \;
\frac{P^{a b}_\mu P^{a b}_\nu}{(P^{a b})^2} \right)
\; \frac{1}{(P^{a b})^2} \; \proj^{a b, c d} \; ,
\label{gluon_prop}
\eeq
These are illustrated in Fig.~\ref{fig_propagators}.

There are several matters of notation to attend to.  All of these
sound more complicated than is true after drawing the corresponding
double line diagram.
In Eqs.~(\ref{quark_momenta}) to (\ref{gluon_prop}), 
we adopt the convention that 
color indices shared between momenta and projection operators are 
{\it not} summed over.

Implicitly, a quark line
carries two arrows: one as a Dirac particle, and one for
color.  Either $P^a_0 = p_0 + Q^a$, if the directions coincide, or
$(P^a)_0 = p_0 - Q^a$, if they do not.  As long as one is consistent with
directions, though, this does not really matter.  
For the quark propagator, this leads to the obvious identity:
\beq
\frac{\delta^{a b}}{-i \! \not \!\!\aPtl +m} =
\frac{\delta^{a b}}{ -i \! \not \!\!\bPtl +m } \; .
\eeq

Similarly, for fields in the adjoint 
representation, where $P^{a b}_0 = p_0 + Q^a - Q^b$, we define the left
index so that it is in the direction of the momentum.  
Thus if we change the direction, $P' = - P$, then we must also reverse the
order of indices, and $(P')^{b a} = - P^{a b}$. For the ghost propagator,
for example,
\beq
\frac{1}{(P^{a b})^2 } \; \proj^{a b, c d}  
= \frac{1}{(P^{d c})^2 } \; \proj^{a b, c d}   \; .
\label{gluon_prop_identity}
\eeq
The same is true for the gluon propagator.  

The vertices between quantum fluctuations, $B_\mu$ and $\psi$,
are also simple, taking care of indices and the like.
The vertex between a quark, antiquark, and a gluon is obtained by taking 
the derivative of the action, ${\cal S}=\int d^{4}x {\cal L}$, as
\beq
-\frac{\delta{\cal S}}{\delta \psi^b (R)\;  \delta B^{d c}_\mu (Q)\;\delta \overline{\psi}^a  (P) }
= i g \; \left(t^{c d}\right)_{a b} \; \gamma_\mu \; .
\eeq
The order of the gluon indices flip, from the left to the right hand side,
because $B_\mu = t^{c d} B^{d c}_\mu$.
The vertex between a ghost, antighost, and a gluon is
\beq
-\frac{\delta{\cal S}}{  \delta \eta^{f e}(R)\;  \delta B^{d c}_\mu(Q) \; \delta \overline{\eta}^{b a}(P) }
= i  g \; f^{( a b , c d , e f)} \; (P^{a b})_\mu \; .
\label{ghost_gluon_vertex}
\eeq
The three gluon vertex is
\beq
-\frac{\delta{\cal S}}{ \delta B^{f e}_\lambda(R)  \;\delta B^{d c}_\nu(Q) \; \delta B^{b a}_\mu(P) }
= - i \, g \; 
f^{( a b , c d, e f)} \; 
\varGamma_{\mu \nu \lambda}(P^{a b},Q^{c d},R^{e f}) \; ,
\eeq
where
\beq
\varGamma_{\mu \nu \lambda}(P^{a b},Q^{c d},R^{e f}) = 
\left(P^{a b}_\lambda - Q^{c d}_\lambda\right) \delta_{\mu \nu} 
+ \left(Q^{c d}_\mu - R^{e f}_\mu\right) \delta_{\nu \lambda} 
+ \left(R^{e f}_\nu - P^{a b}_\nu\right) \delta_{\lambda \mu} 
\; .
\label{three_gluon}
\eeq
These are all the usual vertices, with the replacement of ordinary
momenta by momenta which carry color indices.  Again, in
Eqs.~(\ref{ghost_gluon_vertex}) and (\ref{three_gluon}), color indices
shared by momenta are not summed over.

Momenta
with colored indices satisfy momentum conservation as usual, so
in Eq.~(\ref{three_gluon}),
\begin{equation}
P^{a b}_\mu + Q^{c d}_\mu + R^{e f}_\mu = 0 \; .
\label{momentum_conservation}
\end{equation}
Now consider one of the external momenta, say $R^{e f}_\lambda$, 
contracted with the three gluon vertex.  This satisfies the identity:
\begin{equation}
R^{e f}_\lambda \; \Gamma_{\mu \nu \lambda}(P^{a b},Q^{c d},R^{e f}) =
\varDelta^{-1}_{\mu \nu}(Q^{c d}) - 
\varDelta^{-1}_{\mu \nu}(P^{a b}) \; ,
\label{three_gluon_identity}
\end{equation}
where $\varDelta^{-1}$ is the transverse piece of the inverse 
gluon propagator:
\begin{equation}
\varDelta^{-1}_{\mu \nu}(P) = \delta_{\mu \nu} P^2 - P_\mu P_\nu \; .
\label{inverse_transverse_propagator}
\end{equation}
This is the same identity as for $Q = 0$,
Eq.~(2.50) of Ref.~\cite{htlB}.  This is useful because as for $Q = 0$
\cite{htlB}, it can be used to show that hard thermal loops 
are independent of the gauge fixing condition.

The four gluon vertex has the usual form, a sum over products of structure
constants:
\beq
\begin{split}
-\frac{\delta{\cal S}}{ \delta B^{h g}_\sigma(S) \; \delta B^{f e}_\lambda(R)  \;\delta B^{d c}_\nu(Q)\;  \delta B^{b a}_\mu(P) }
=& - g^2 \sum_{i,j=1}^\Nc \left( 
f^{( a b , c d, i j)} f^{( e f , g h, j i)} 
\left( \delta_{\mu \lambda} \delta_{\nu \sigma} 
- \delta_{\mu \sigma} \delta_{\nu \lambda}  \right)
\right. \\
&\qquad\qquad
+ f^{( a b , e f, i j)} f^{( g h , c d, j i)} 
\left( \delta_{\mu \sigma} \delta_{\lambda \nu} 
- \delta_{\mu \nu} \delta_{\lambda \sigma}  \right) \\
&\qquad\qquad+\left. f^{( a b , g h, i j)} f^{( c d , e f, j i)} 
\left( \delta_{\mu \nu} \delta_{\sigma \lambda} 
- \delta_{\mu \lambda} \delta_{\sigma \nu}  \right)
\right)
\; .
\end{split}
\label{four_gluon}
\eeq

We conclude by noting that Eq.~(\ref{gluon_prop_identity}) can be used
to simplify insertions of ghost or gluon lines in loop diagrams.  
If a gluon ties onto a quark line, then in Feynman gauge this enters as
\begin{equation}
t^{b a} \; \frac{\proj^{a b,c d} }{(P^{a b})^2} \; t^{d c} \; ,
\label{insert_gluon_quark_line}
\end{equation}
where we neglect the rest of the diagram.
The projection operator in the gluon propagator is a sum of two terms,
Eq.~(\ref{projector}).  
Since the gluon appears inside the loop, the $c$ and $d$ indices are
summed over.  
Unlike the $a$ and $b$ indices, which also enter through $P^{a b}$,
this is the {\it only} place where $c$ and $d$ indices enter.
Since the generators are traceless, though, Eq.~(\ref{sum}), 
any contribution from the second term 
in the projection operator, $- \delta^{a b} \; \delta^{c d} /\Nc$, vanishes.  Hence
in the gluon line, we can replace the projection operator by the first term,
which is just a Kronecker delta, 
$\proj^{a b, c d} \rightarrow \delta^{a d}  \delta^{b c}$, so 
Eq.~(\ref{insert_gluon_quark_line}) becomes
\begin{equation}
t^{b a} \; \frac{1}{(P^{a b})^2} \; t^{a b} \; .
\label{insert_gluon_quark_line_simplify}
\end{equation}
The same is true in any gauge.  It is also true for a gluon tied to
either a three gluon or four gluon vertex, since in each case, 
Eqs.~(\ref{three_gluon}) and (\ref{four_gluon}),
the projection operator in the gluon propagator ties onto a factor
$f^{(c d,ef,gh)}$, and that any such structure constant involves
a commutator of $t^{c d}$, Eq.~(\ref{structure_const}).
The same holds for a ghost propagator tied onto a ghost antighost gluon
vertex.  This useful simplification was first seen in Eq.~(\ref{sum_squared_gens})
in Sec.~\ref{sec_basis}: generators can be contracted not
with projection operators, but just with ordinary Kronecker deltas.

\subsection{Propagators in a mixed representation}
\label{sec_mixed_props}

In this section we discuss a useful trick for computing scattering amplitudes,
starting in the imaginary time.  To avoid unnecessary complication,
we replace the color matrices, either $Q^a$ or $Q^{a b}$, by a single
background field, $Q$.  This is identical to considering the propagation
of an electron in QCD, in the presence of a background field $A_0 \sim Q/e $.
The extension to QCD is automatic, as will be clear from the examples
which follow in later sections.

We then introduce a ``mixed'' representation for the propagators
\cite{htlA,htlB,lebellac}.  For the spatial directions, one works as usual
in momentum space, but for the time direction, instead one stays in coordinate
space.  For example, consider a propagator, 
\beq
\varDelta_Q(\tau,E) = T \; \sum_{n= -\infty}^{+ \infty}
\; \frac{{\rm e}^{- i ( p_0 + Q) \tau}}{(p_0 + Q)^2 + E^2} \; ;
\label{prop1}
\eeq
$E$ is the energy, typically $E = \sqrt{\vec{p}^{\; 2} + m^2}$.
We assume the field is bosonic, so the Euclidean energy
$p_0 = 2 \pi n T$, for integral $n$.  
For the time being, we also assume that $0 \leq \tau \leq 1/T$.
The sum is performed by contour integration in the complex
$p_0$ plane.  There are two poles, for $p_0 = - Q \pm i E$, which give
\beq
\varDelta_Q(\tau,E) = \sum_{s = \pm} \frac{s}{2 E} \left(
1 + n(s E - i Q) ) \right) \; {\rm e}^{- s E \tau} \; .
\label{prop2}
\eeq
Here $n(E)$ is the usual Bose-Einstein statistical distribution function,
\beq
n(E) = \frac{1}{{\rm e}^{E/T} - 1} \; ,
\label{boseeinstein}
\eeq
so the only change for $Q \neq 0$ is the change in the
statistical distribution function,
\beq
n(E \mp i Q) = \frac{1}{{\rm e}^{(E\, \mp \, i Q)/T} - 1} \; .
\label{boseeinsteinQ}
\eeq
Notice that this is the only place where $Q$ enters into Eq.~(\ref{prop2}):
the propagators in Euclidean time, $\sim \exp({\mp E\tau})/2E$, are
{\it identical} to that for $Q=0$.

This can be understood by recognizing
that the parameter $i Q$ enters exactly like 
a chemical potential, albeit one which is imaginary.
Because of this, the associated statistical distribution functions, 
$n(E \mp i Q)$, are complex valued.  
It helps to rewrite Eq.~(\ref{prop2}) in a less compact form,
\beq
\varDelta_Q(\tau,E) = \frac{1}{2 E} \left(
( 1 + n(E-iQ) ) \; {\rm e}^{- E \tau}
+ n(E+i Q) \; {\rm e}^{+ E \tau} \right ) \; .
\label{prop3}
\eeq
This form is physically more transparent.  
The first term, with propagator ${\rm e}^{ - E \tau}/2E$,
is proportional to $1+ n(E-iQ)$.
The $1$ is the contribution in vacuum, while $n(E-iQ)$ 
represents the induced emission of a particle, with energy $E$ and 
chemical potential $+iQ$, into the thermal bath. The second term, 
with propagator $\exp(+ E \tau)/2E$, is proportional to
$\sim n(E+iQ)$.  This represents absorption of a field with energy $E$,
and chemical potential for the antiparticle, 
$-iQ$, from the thermal bath.  

This expression for the propagator is the same as for a real chemical
potential, $\mu$, except that $\mu$ is replaced by $i Q$.
As when $\mu \neq 0$, the statistical distribution
functions are modified, but the energies of the system
remain unchanged.  The same is then true for an imaginary chemical
potential, $iQ$.  This is why the form of the propagators in
imaginary time, $\exp({\pm E \tau})/2E$, are unaffected by $Q$.
We argue in the next subsection that this remains valid for propagation
in real time as well.

We make some remarks to help illuminate the meaning of the statistical
distribution functions when $0 < Q < \pi T$.
First, when $Q=0$, the Bose-Einstein distribution function
is singular as $E \rightarrow 0$, $n(E) \sim T/E$.  This
singularity is related to the phenomenon of 
Bose-Einstein condensation at low
temperature.  In contrast, whenever $Q \neq 0$, 
the distribution function is regular as the energy vanishes,
$n(0-iQ) = 1/(\exp(-iQ/T) - 1)$.   This includes ordinary
fermions, when $Q = \pi T$, and $n(-i \pi T) = -1/2$.

It is also helpful to consider adding a real
chemical potential, $\mu$, in addition to $i Q$.
For ordinary fermions, $Q = \pi T$, the Fermi-Dirac distribution function
with $\mu \neq 0$ is
\begin{equation}
\widetilde{n}(E-\mu) = - n(E - i \pi T - \mu) 
= \frac{1}{{\rm e}^{(E-\mu)/T} + 1} \; .
\label{fermi_sea}
\end{equation}
In the limit of zero temperature, 
if $E > \mu$, $\widetilde{n} = 0$, while 
if $E < \mu$, $\widetilde{n} = 1$.  For antiparticles,
$\widetilde{n}(E +\mu)=0$ for any $E$ as $T \rightarrow 0$. 
This is just a Fermi sea, and represents a net excess of particles
over antiparticles.

When $Q \neq \pi T$, if there is also 
a real chemical potential, $\mu$, the associated
statistical distribution function is
\begin{equation}
n(E-\mu- i Q) = \frac{1}{{\rm e}^{(E-\mu-iQ)/T} - 1} \; .
\end{equation}
Taking $Q = 2 \pi T q$, as $T \rightarrow 0$, 
if $E > \mu$, $n = 0$.  However,
if $E < \mu$, $n = - 1$; the negative sign of $n$ is natural,
see Eq.~(\ref{fermi_sea}).
For antiparticles, $n(E + \mu + i Q)$, one
finds $n = 0 $ for all energies.  Thus a real chemical potential introduces
an asymmetry between particles and antiparticles for all $Q \neq 0$.

Speaking loosely, when $0 < Q < \pi T$ 
particles behave with something like fractional statistics.  
This analogy is not precise, though, 
merely suggestive.  In particular, for both
cases in which $Q$ arises, a given $Q$ is not physical.  For a $\ZN$
interface, $Q$ is a function of $z$, and one integrates over all $Q(z)$.
In the semi-QGP, there is a distribution of $Q$'s, and it is only 
integrals over the distribution which are physically meaningful.  In both
cases, after summing over all $Q$'s, the usual relationship between spin
and statistics is recovered.  

We conclude with some useful identities.  The first is 
\begin{equation}
1 + n(E - i Q)  = {\rm e}^{(E - i Q)/T} \; n(E- i Q) \; .
\label{statistical_identity_A}
\end{equation}
This is well known for $Q=0$, and by construction, must then be true
for $Q \neq 0$, by simply replacing $E \rightarrow E - i Q$.  
It applies for either sign of $Q$.

The propagator in Eq.~(\ref{prop1}) is defined for positive $\tau$,
$0 \leq \tau \leq 1/T$.  The extension to negative values,
$- 1/T \leq \tau \leq 1/T$, is
\beq
\varDelta_Q(\tau,E) = \sum_{s = \pm} \frac{s}{2 E} \left(
1 + n(s E - i Q \; {\rm sign}(\tau)\, ) \, 
\right) \; {\rm e}^{- s E |\tau|} \; .
\label{prop4}
\eeq
From this, or directly from Eq.~(\ref{prop1}), 
\beq
\varDelta_{Q}(-\tau,E) = \varDelta_{-Q}(\tau,E) \; .
\label{prop5}
\eeq
From this it also follows that
\beq
\varDelta_Q(\tau - 1/T,E) = {\rm e}^{i Q/T} \; \varDelta_{Q}(\tau,E) \; .
\label{kms}
\eeq
This is the generalization of the Kubo-Martin-Schwinger condition
\cite{lebellac} to a background field $Q$.

In practice we will start with diagrams in momentum space, and
then transform a sum over $p_0$ to an integrals over
$\tau$'s.  This is done by using
the Fourier transform of Eq.~(\ref{prop1}), which is
\beq
\frac{1}{(p_0 + Q)^2 + E^2} =
\int^{1/T}_0 \; d\tau \; \frac{e^{i (p_0 + Q) \tau}}{2 E}
\left( \left( 1 + n(E - i Q) \right) \rme^{- E \tau}
+ n(E + i Q) \rme^{+ E \tau} \right) \; .
\label{prop_inv}
\eeq

In computation we also require 
\beq
\frac{p_0 + Q}{(p_0 + Q)^2 + E^2} =
\int^{1/T}_0 \; d\tau \; \frac{-i}{2 E}
\left( \frac{\partial}{\partial \tau} e^{i (p_0 + Q) \tau}
\right) 
\left( \left( 1 + n(E - i Q) \right) \rme^{- E \tau}
+ n(E + i Q) \rme^{+ E \tau} \right) \; .
\label{prop_inv_B}
\eeq
Integrating by parts, this equals
\beq
 -i \left(
\rme^{i(p_0 + Q)/T} \varDelta_Q(1/T,E) - \varDelta_Q(0,E) \right)
+ \int^{1/T}_0 \; d\tau \rme^{i(p_{0}+Q)\tau} 
\; i \frac{\partial}{\partial \tau} \varDelta_Q(\tau,E) \; .
\label{prop_inv_C}
\eeq
The first term vanishes, since $p_0$ is bosonic, so $\rme^{i p_0/T} = 1$,
and by the condition of Eq.~(\ref{kms}), for
$\tau = 1/T$.  The second term is easy to evaluate, and gives
\begin{equation}
\frac{p_0 + Q}{(p_0 + Q)^2 + E^2} =
\int^{1/T}_0 d\tau \;
\frac{e^{i (p_0 + Q) \tau}}{2 E} 
\left( \left( 1 + n(E - i Q) \right) (- i E) \rme^{- E \tau}
+ n(E + i Q) (+ i E) \rme^{+ E \tau} \right) \; .
\label{prop_inv_D}
\end{equation}
Integrals with higher powers of $p_0 + Q$ are not required, since they
can be reduced as
\begin{equation}
\frac{(p_0 + Q)^2}{(p_0 + Q)^2 + E^2}
= 1 - \frac{E^2}{(p_0 + Q)^2 + E^2} \; ,
\end{equation}
which can be handled by previous results.
\subsection{Amplitudes in real time}
\label{sec_analytic_cont}

In this section we follow Furuuchi \cite{furuuchi}
and discuss how to proceed from amplitudes, 
computed in imaginary time, to scattering amplitudes.

We remark that Smilga \cite{smilga} has argued that $\ZN$ domain walls
are entirely a construction valid only in imaginary time, and have
no relevance for scattering amplitudes.  If $\ZN$ domain structure is a
natural consequence of the system in thermal equilibrium ---
i.e., in imaginary time --- then it is difficult to see how
they could not be relevant for small fluctuations about thermal 
equilibrium, which is what amplitudes in real time represent.  
Notably, lattice simulations of the dynamical evolution of $\ZN$
domains have been performed \cite{real_time_lattice}, and as one
might expect, are very
similar to the evolution in Potts models, to which they are closely 
analogous.  We do acknowledge, however, that the lattice time step is not
immediately related to a physical time.

\begin{figure}
\includegraphics[width=0.5\textwidth]{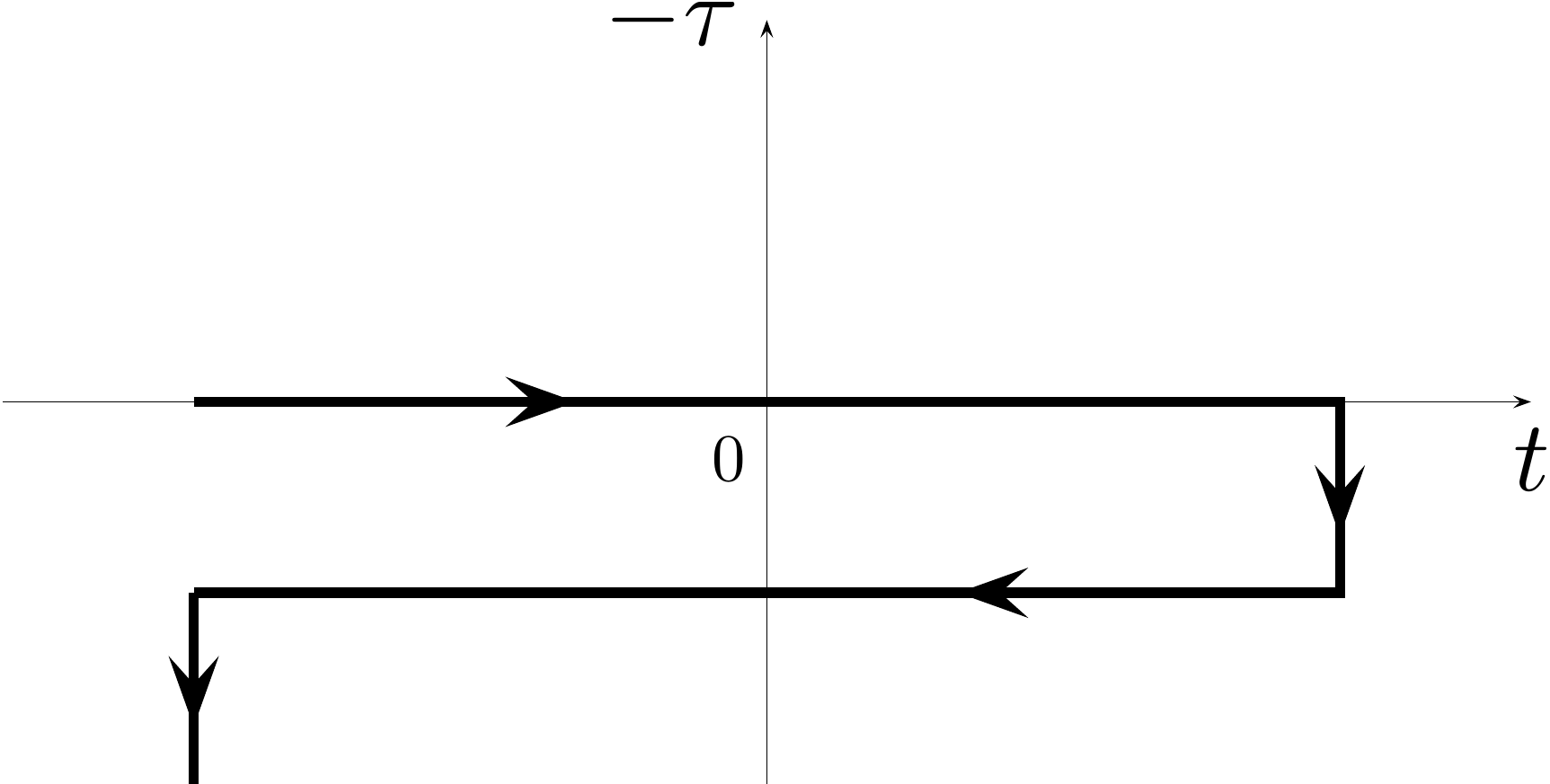}
\caption{Contour in real and imaginary time.}
\label{time_contour}
\end{figure}

Consider the usual contour in the plane of real and imaginary time.
The imaginary time variable $\tau$ runs from $0$ to $1/T$, and represents
a thermal ensemble in thermal equilibrium.  The contour also runs
in real time, $t$, from $0$ to $\infty$, and then back again,
representing fluctuations about the thermal ensemble.
This is illustrated in a standard figure, Fig.~\ref{time_contour}.
The exact shape of the contour \cite{lebellac} will not matter for our
purposes.

The essential question is on what parts of the contour the background field
$A_0$ is nonzero.  The answer is to take the background $A_0$ field
{\it only} for the part of the contour in imaginary time, and {\it not}
for the part of the contour in real time.  
This is absolutely necessary for the integrals to be well defined.  Since
the real time runs from $-\infty$ to $+\infty$, and then back again, if
there was such a background field, it would affect the behavior at large
times.

This is clearest in considering the background $Q$ field as a chemical
potential for color charge \cite{imag_chem_pot}.  
A real chemical potential alters the initial 
statistical distribution of the particles: for fermions, for example,
it represents a net excess of particles over antiparticles, or a Fermi
sea.  While the canonical momenta are shifted by a chemical potential,
the evolution in real time is by the usual Hamiltonian of the system.
Thus the evolution, in real time $t$, of some operator $\phi$ proceeds by
the usual Heisenberg relation, $\phi(t) = {\rm e}^{+ i H t} \phi(0)
{\rm e}^{- i H t}$.

While
the $Q$ field represents an imaginary chemical potential, its effect
is only to alter the initial color distribution of particles.  
The canonical momentum is shifted, but not the Hamiltonian.  

In practice, one computes
an amplitude with $Q$ dependent momenta: $p_0^{a b}$ for gluons,
$p_0^a$ for quarks.  The above implies that one amplitudes are constructed
by taking
\beq
p^{ab}_0 = p_0 +  (Q^a - Q^b) \rightarrow - i \omega^{a b} \; ,
\label{analytic_cont}
\eeq
for gluons, and similarly for quarks.  Here, $\omega$ is an energy in
real time, and as such, can take arbitrary values.

The division into hard momenta, of order $\sim T$, and soft momenta,
$\sim gT$, is preserved by this procedure.  The usual Euclidean $p_0$
is a multiple of $2 \pi T$ for bosons.  Thus, the shift by the fractional
amount, $\sim Q$, does not change this.  In contrast, the Minkowski energy
$\omega^{a b}$ is a continuous variable, and it is consistent to assume that
it is soft.  

There is an important subtlety which we ignore.  Usual scattering 
amplitudes are invariant under arbitrary reparametrizations of the fields.
It is far from clear that this is true for scattering in a {\it fixed}
$Q$ field.  The point is that it is necessary to compute a physical
process.  For a $\ZN$ interface, this would be scattering not at a given
point, but integrated over the entire spatial extent of the interface.  
Similarly, for the semi-QGP, scattering in a fixed $Q$ field could well
exhibit unphysical behavior.  The physical quantity there are amplitudes
in which one integrates over the entire distribution of $Q$'s, representing
the thermal equilibrium state.

\section{Quark Self-Energy}
\label{sec_quark_self_energy}

With the formalism in place, we proceed to computing the self-energy for
a quark in a background field, $Q \neq 0$.  The computations are
relatively straightforward, and do not exhibit complications which
will arise for the gluon self-energy in the next section.  We go through
this example in some detail, so 
that the reader can develop familiarity with computing in the
presence of a background field.
It also helps to understand the novelty of the new terms in the gluon
self-energy.

At one-loop order, the standard diagram is, in our notation,
\begin{equation}
-\varSigma(\aPtl)_{a b}=
-\, g^2\; (t^{d e})_{a c}\; {\cal P}_{d e,f g}
\; (t^{f g}) _{c b}\;
\TInt\;
\frac{\gamma_\mu\, i\otherSlash\cKtl\, \gamma^{\mu}}
{(\aPtl-\cKtl)^2 (\cKtl)^2} \; .
\label{qk_self_1}
\end{equation}
Here $\cKtl$ 
is the momentum of the quark in the loop, 
and $\aPtl -\cKtl$ the momentum of the
gluon.  Thus $\cKtl$ is a fermionic momentum, and
$\aPtl- \cKtl = P^a - K^c$ is bosonic.

In Eq.~(\ref{qk_self_1}) the integral is that appropriate for
a bosonic field at nonzero temperature,
\begin{equation}
\TInt = \TTInt \; \; , \;\; k_0 = 2 \pi n T \; .
\label{qk_self_1A}
\end{equation}
Remember that we can take $p_0$ and $k_0$ to be bosonic, by using
a background field which is $\aQtl = Q^a + \pi T$.

\begin{figure}
\includegraphics[width=0.75\textwidth]{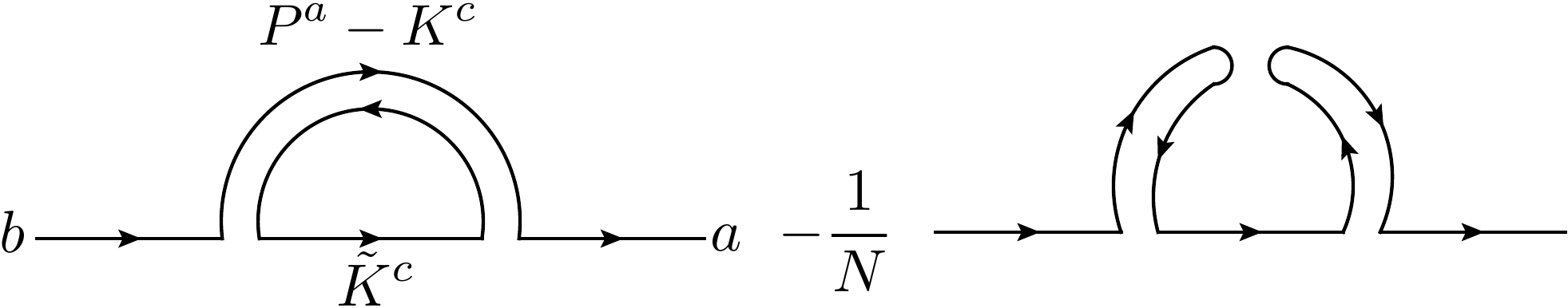}
\caption{One-loop diagram for the quark self-energy.}
\label{fig_quark_self_energy}
\end{figure}

The color structure reduces immediately.  
By Eqs.~(\ref{insert_gluon_quark_line}) and 
(\ref{insert_gluon_quark_line_simplify}), 
the gluon projection operator can be replaced by an ordinary Kronecker delta
\begin{equation}
-\varSigma(\aPtl)_{a b}= \, ig^2\; {\cal P}_{a c ,c b}\;
\TInt
\frac{\otherSlash\cKtl }{(P^a - K^c)^2(\cKtl)^2}
 \; .
\label{qk_self_2}
\end{equation}
The color structure is illustrated by
the diagram of Fig.~\ref{fig_quark_self_energy}.
This is a sum of the planar diagram, minus $1/\Nc$ times a diagram
in which all indices are equal.  

We wish to extract the hard thermal loop, $\sim T^2$, from 
Eq.~(\ref{qk_self_2}).  Instead, to simplify the discussion, we consider
the integral
\begin{equation}
{\cal I}(\aPtl) = \TInt \; 
\frac{1}{(P^a - K^c)^2(\cKtl)^2} \; ,
\label{scalar_qk_self_energy}
\end{equation}
in full generality.  

To perform the sum over $n$, it is useful to use the mixed representation,
Sec.~\ref{sec_mixed_props}.  Using Eq.~(\ref{prop_inv}), we write
\begin{equation}
\frac{1}{(\cKtl\, )^2}
= \int^{1/T}_0 \; d\tau \; \frac{ \rme^{i (k_0 + \cQtl) \tau}}{2 E_k}\;
\left( ( 1 + n(E_k - i \cQtl)) \rme^{-E_k \tau}
+ n(E_k + i \cQtl) \rme^{+ E_k \tau} \right) \; .
\label{qk_prop_1}
\end{equation}
for the quark like propagator, $E_k = \sqrt{\vec{k}^2}$, and
\begin{equation}
\begin{split}
\frac{1}{(P^a - K^c)^2}
= \int^{1/T}_0 \; d\tau' \; 
\frac{ \rme^{i (p_0 -k_0 + Q^{a c}) \tau'}}{2 E_{p-k}}\;
\left( ( 1 + n(E_{p-k} - i Q^{a c})) \rme^{-E_{p-k} \tau'}
\right. \\
\left.
\qquad+ \; n(E_{p-k} + i Q^{a c}) \rme^{+ E_{p-k} \tau'} \right) \; .
\end{split}
\label{qk_prop_2}
\end{equation}
for the gluon propagator, $E_{p - k} = \sqrt{(\vec{p}-\vec{k})^2}$.

The sum over $n$, where $k_0 = 2\pi n T$, can then be performed immediately,
and gives a delta function in time,
\begin{equation}
T \sum_{-\infty}^{+\infty} {\rm e}^{i k_0(\tau - \tau')}
= \delta(\tau - \tau') \; .
\label{delta_function}
\end{equation}
We divide this up into four integrals,
\begin{equation}
{\cal I} = \int \frac{d^3 k}{(2 \pi)^3} \; \frac{1}{(2 E_k)(2 E_{p-k})}
\left( {\cal I}_1 + {\cal I}_2 + {\cal I}_3 + {\cal I}_4 \right) \; .
\label{sum_integrals}
\end{equation}
All of the terms are integrals over $\tau$.  The first is
\begin{equation}
{\cal I}_1(\cQtl,Q^{ac})
= \int^{1/T}_0 d \tau
\; {\rm e}^{(ip_0 + i Q^{ac}+ i \cQtl - E_k - E_{p-k})\tau}
(1 + n(E_{k}-i\cQtl))(1+n(E_{p-k}-iQ^{a c})) \; ,
\label{first_integral}
\end{equation}
with the other three of a similar form.  The integral is easy to do,
\begin{equation}
{\cal I}_1(\cQtl,Q^{ac}) = 
\; \left( \frac{{\rm e}^{(ip_0 + iQ^{ac}+i\cQtl - E_k - E_{p-k})/T}
- 1}{ip_0 + i \aQtl - E_k - E_{p-k}}\right)
(1 + n(E_{k}-i\cQtl))(1+n(E_{p-k}-iQ^{a c})) \; .
\label{first_integral_B}
\end{equation}
In the energy denominator, we rewrite
\begin{equation}
Q^{a c} + \cQtl= Q^a - Q^c + Q^c + \pi T = 
 \aQtl \; .
\label{one_Q_identity}
\end{equation}
For the other terms, though, it is better
not to use this.  Since $p_0$ is a bosonic momentum, 
$\exp({i p_0/T}) = 1$, and we group 
\begin{equation}
{\rm e}^{(ip_0 + i Q^{ac}+i\cQtl - E_k - E_{p-k})/T}
= \rme^{-(E_k - i\cQtl)/T} \;
\rme^{-(E_{p-k} - i Q^{a c})/T} \; .
\label{first_integral_C}
\end{equation}
This is useful because the identity of Eq.~(\ref{statistical_identity_A})
can now be brought to bear, so that
\begin{equation}
\begin{split}
{\cal I}_1 (\cQtl,Q^{ac})= &
\; \frac{ n(E_k-i\cQtl)n(E_{p-k}-iQ^{a c}) -(1 + n(E_k-i\cQtl))
(1+n(E_{p-k}-iQ^{a c})) }{ip_0 + i \aQtl - E_k - E_{p-k}} \\
= & \; \frac{ -1 }{i\aptl_0 - E_k - E_{p-k}} \; 
\left( 1 + n(E_k-i\cQtl) +n(E_{p-k}-iQ^{a c}) \right) 
\; . 
\end{split}
\label{first_integral_D}
\end{equation}
This is the exact same expression as for $Q=0$, with the replacement 
\begin{equation}
E_k \rightarrow E_k - i \cQtl \;\; ; \;\;
E_{p-k} \rightarrow E_{p-k} - i Q^{a c} \; .
\label{shift_energies}
\end{equation}
This holds for the energy denominator as well: that for $Q=0$,
$ip_0 - E_k - E_{p - k}$, becomes $i\aptl_0 - E_k - E_{p - k}$
after the shift of Eq.~(\ref{shift_energies}), 
using Eq.~(\ref{one_Q_identity}).  Note, however, that this substitution
is not universal, and holds only for the ${\cal I}$'s: 
in Eq.~(\ref{sum_integrals}),
the residues for the propagators remain $1/(2 E_k)$ and $1/(2 E_{p-k})$,
respectively.  

We went through this derivation
in detail, because 
the chemical potentials are imaginary, so that some care is in order.
Even so, the manipulations for $Q \neq 0$ are 
very similar to those for $Q=0$.
For example, in going from the first line in Eq.~(\ref{first_integral_D})
to the second line, that terms involving two
statistical distribution functions,
$n(E_k-i\cQtl) n(E_{p-k}-iQ^{a c})$, drop out is just the usual
cancellation between stimulated emission and absorption in a thermal
bath.  

The other integrals can be done similarly, and follow from the result
for $Q=0$, by shifting the energies, as in Eq.~(\ref{shift_energies}),
using the identity of Eq.~(\ref{one_Q_identity}):
\begin{align}
{\cal I}_2(\cQtl,Q^{ac}) = &\; \frac{ 1 }{i\aptl_0 - E_k + E_{p-k}} \; 
\left( n(E_k-i\cQtl)  - n(E_{p-k} + iQ^{a c}) \right) 
\; ,
\label{first_integral_F} \\
{\cal I}_3 (\cQtl,Q^{ac})= &\; \frac{ -1 }{i\aptl_0 + E_k - E_{p-k}} \; 
\left( n(E_k + i\cQtl) - n(E_{p-k}-iQ^{a c}) \right) 
\; ,
\label{first_integral_G}\\
{\cal I}_4 (\cQtl,Q^{ac})=& \; \frac{ 1 }{i\aptl_0 + E_k + E_{p-k}} \; 
\left( 1 + n(E_k+i\cQtl) +n(E_{p-k}+iQ^{a c}) \right) 
\; . 
\label{first_integral_H}
\end{align}

The extension to the original integral of Eq.~(\ref{qk_self_2}) is
immediate.  The term $\sim \vec{k} \cdot \vec{\gamma}$ is the same
as above.  That $\sim \cktl_0\gamma_0$ is evaluated using the identity
of Eq.~(\ref{prop_inv_D}).  In this context, this tells us to
replace 
$\cktl_0 \rightarrow \mp i E_k$.
For the terms with positive energy, ${\cal I}_1$ and ${\cal I}_2$, 
$- i E_k$ enters; for those with negative energy, 
${\cal I}_3$ and ${\cal I}_4$, $+ i E_k$.  

We next extract the hard thermal loop from Eq.~(\ref{qk_self_2}).
These are terms where the external momenta are soft,
$p \sim g T$, and are as large as the corresponding term
at tree level. 
In the quark self-energy, then, the hard thermal
loops are $\sim g^2 T^2/p$.

The loop momenta in a hard thermal loop are hard, $k \sim T$.
In the statistical distribution
functions we can then approximate $E_{p - k} \sim E_k = k$.  
The dominant terms arise from the energy denominators with Landau damping,
${\cal I}_2$ and ${\cal I}_3$.  These terms are dominant because the
energy denominators are a difference of large energies, and thus are small:
\begin{equation}
i\aptl_0 \pm (E_k - E_{p-k})
\approx i \aptl_0 \pm p \cos \theta \; ,
\label{energy_denominator_reduction}
\end{equation}
where $\cos \theta = \hat{k} \cdot \hat{p}$.  

Notice that these terms are only dominant if --- and only if ---
the energy $i \aptl_0$ is small after analytic continuation,
Sec.~\ref{sec_analytic_cont}.  
If only $i p_0$, and not $i \aptl_0$, were small after analytic
continuation, then $i \aQtl$ would be a hard momentum, and
there would be no hard thermal loop in the diagram: everything would
be a correction which is suppressed by at least $\sim g$ relative to the 
propagator at tree level.

Introducing the vector $\hat{K} = (i, \hat{k})$, 
$\hat{k}^2 = 1$, the hard thermal loop
in the quark self-energy becomes
\begin{equation}
-\varSigma(\widetilde{P}^{a})_{a b}\HTLapprox
(m^2_\text{qk})_{a b} \; \delta\varSigma(\aPtl) \; ,
\label{htl_quark_A}
\end{equation}
where the $\HTLapprox$ sign indicates that the hard thermal loops are equal,
but not (necessarily) terms beyond that order, and 
\begin{equation}
\delta\varSigma(P) \; = \; \int
\frac{d \Omega}{4 \pi} \; \frac{i\Slash{\hat{K}}}{P \cdot \hat{K}} \; ;
\label{htl_quark_A1}
\end{equation}
the angular integral is over all directions of the unit vector $\hat{k}$.
The function $\delta\varSigma(P)$ is identical to that for $Q = 0$.
In the quark self-energy, this function is multiplied by a thermal
quark ``mass'', which is a function of $Q$:
$$
(m^2_\text{qk}(Q))_{a b} = \frac{g^2 T^2}{24} \; 
\sum_{c=1}^\Nc \; {\cal P}_{a c ,c b}\; 
\left( \Acal(Q^{a c}) - \Acal(\cQtl) \right) 
$$
\begin{equation}
= \delta_{a b} \; \frac{g^2 T^2}{24} \; 
\left( \sum_{c=1}^\Nc \left( \Acal(Q^{a c}) - \Acal(\cQtl) \right)
- \frac{1}{\Nc} \left( \Acal(0) - \Acal(\aQtl) \right) \right) \; ,
\label{htl_quark_A2}
\end{equation}
where
\begin{equation}
\Acal(Q) = 
\frac{3}{\pi^2 T^2} \int_0^\infty \; d k \; k\; \left(n(k - i Q)
+ n(k + i Q) \right) \; .
\label{htl_quark_B}
\end{equation}
We normalize $\Acal(Q)$ in anticipation of the final result.
For ordinary hard thermal loops, the integrals over the hard, loop momenta
decouple into an angular integral times an integral over the 
statistical distribution functions.  Equation~(\ref{htl_quark_A}) shows that
this remains true for the quark hard thermal loop when $Q \neq 0$.  
The same is also true for the gluon hard thermal loop.

For the quark self-energy, in the end the color structure is no different
than the propagator at tree level, $\sim \delta_{a b}$.  
New color structures do arise for the hard thermal loop in the gluon
self-energy.

The integrals over statistical distribution functions when $Q \neq 0$
are not much more difficult than for $Q=0$ \cite{interface1}.
We write
\begin{equation}
Q = 2 \pi T q \; \, .
\end{equation}
Then
$$
\frac{6}{\pi^2 T^2}\; \int_0^\infty \; dk \; k\;
n(k - i Q)
= \frac{6}{\pi^2} \int^\infty_0 dk \; k \;
\frac{\rme^{-k + 2 \pi i q}}{1 - \rme^{- k + 2 \pi i q}} 
= \frac{6}{\pi^2} \; \sum_{j=1}^\infty
\frac{1}{j^2} \; \rme^{2 \pi i q j}
\; 
$$
\begin{equation}
= \; 
1 - 6 q(1-q) + 
i \,\mathrm{Cl}_{2}(2\pi q) \;,
\label{sum_dist_fncs}
\end{equation}
where $\mathrm{Cl}_{n}(\theta)$ is the Clausen function.
As is typical of similar expressions for a $\ZN$ interface, this is
valid only for $0<q < 1$, but the extension to other values is direct.
The imaginary term $\mathrm{Cl}_{2}(2\pi q)$ 
cancels in the sum which
enters into $\Acal$, so that
\begin{equation}
\Acal(Q) = 1 - 6 q (1 - q) \; .
\label{function_A_Q}
\end{equation}

The hard thermal loop in the quark self-energy is identical to that
for $Q = 0$, up to the change in the thermal quark mass.
To check that one obtains the usual value for $Q=0$, remember
that when all $Q=0$, $\cQtl = \pi T$, or $q = 1/2$.  Since
$\Acal(0) = 1$ and $\Acal(1/2) = - 1/2$, 
\begin{equation}
m^2_{qk}(0)_{a b} = 
\frac{\Nc^2 - 1}{2 \Nc} \; \frac{g^2 T^2}{8} \; \delta_{a b} \; .
\label{usual_quark_mass}
\end{equation}

There is a simple interpretation of the function $\Acal(Q)$.  
Although classically there is no potential for $Q$, 
in the presence of a background field $Q$, a potential is generated
at one-loop order.  For convenience we normalize this potential as
\begin{equation}
{\cal V}(Q) = \frac{1}{2} \; q^2 (1 - q)^2 \; .
\label{define_one_loop_potA}
\end{equation}
Then
\begin{equation}
\Acal(Q) = \; \frac{d^2}{d q^2} \; {\cal V}(Q) \; .
\label{define_one_loop_potB}
\end{equation}
Hence the thermal quark mass is naturally the second derivative of a
potential, as one would expect.

\section{Gluon self-energy}
\label{sec_gluon_self_energy}
\subsection{Gluonic hard thermal loops: tadpoles}
\label{subsec_gluon_htls_tadpoles}

From the example of the quark self-energy, one might expect that the
hard thermal loop in the gluon self-energy when $Q \neq 0$ is just like
that when $Q = 0$, with the same functional form, and the only change
a relatively trivial redefinition of the thermal mass.

We shall see that this is not true: there are new terms which arise
uniquely for non-Abelian gauge fields.  These are not present either
for fermions, coupled to either Abelian or non-Abelian gauge fields, nor
for Abelian gauge fields.

Before delving into the details of the computation in a non-Abelian
gauge theory, in the next two subsections we discuss the differences
between hard thermal loops in the gluonic self-energy when $Q \neq 0$,
versus $Q = 0$.  Hopefully this will make the origin of the new terms
less obscure.  

Hard thermal loops are one-loop diagrams which are as large as the
corresponding terms at tree level when the external 
momenta are soft, $P \sim g T$.  For the gluon self-energy in zero
field, at tree level the inverse propagator is $\sim P^2 \sim (g T)^2$
for soft $P$.  Thus the hard thermal loops are diagrams which are
$\sim g^2 T^2$ times a dimensionless function of the energy, $ip_0 =  \omega$,
divided by the spatial momentum, $p$.  

The simplest hard thermal loop is present in (massless) scalar field theories,
and is just a tadpole integral,
\beq
\TInt \; \frac{1}{K^2} = \frac{T^2}{12} \; .
\label{tadpole_A}
\eeq

Consider the extension of this integral to $Q \neq 0$:
\beq
\TInt \; \frac{1}{(k_0 + Q)^2 + \vec{k}^{\; 2}} \; .
\label{tadpole_B}
\eeq
We use the representation of the propagator in Eq.~(\ref{prop_inv_D}).
The sum over $k_0$ generates $\delta(\tau)$, so the $\tau$ integral
is trivial, and we are left with a single integral over 
$k = \sqrt{\vec{k}^{\;2}}$, so that Eq.~(\ref{tadpole_B}) becomes
\beq
 \frac{1}{4 \pi ^2} \; 
\int_0^\infty dk \; k \; \left( n(k - i Q) + n(k + i Q) \right) 
= \frac{T^2}{12} \; \Acal(Q) \; .
\label{tadpole_C}
\eeq
By comparison with Eq.~(\ref{htl_quark_B}), this is the function $\Acal(Q)$
we found for the quark self-energy, Eq.~(\ref{function_A_Q}).
(An ultraviolet divergent term at zero temperature,
$\sim \int k \; dk $ has been dropped.  
For a scalar field, this is part of a mass divergence.  For a gauge
field, the sum of all such
terms vanishes at zero temperature by gauge invariance.)

A less trivial example is given by the integral
\beq
\TInt \frac{k_0 + Q}{(k_0+Q)^2 + \vec{k}^{\;2}} 
= \; \frac{\pi T^3}{3} \nAcal(Q) 
\; .
\label{scalar_k0_A}
\eeq
As we shall see in Sec.~\ref{sec_one_point_gluon}, 
there is a contribution to the
{\it one} point function of the gluon from such a term, where $k_0 + Q$ arises
from the three gluon vertex.  We stress that such a diagram does {\it not}
arise for the two point gluon function.

This integral vanishes when $Q=0$ because it is odd in $k_0$,
and one sums over both negative and positive values of $k_0$.  Using
the same tricks as above, 
\beq
\nAcal(Q)
= \frac{3}{4i \pi^3 T^3} \; 
\int^\infty_0 dk \; k^2 \left( n(k - i Q) - n(k + i Q) \right) \; .
\label{scalar_k0_B}
\eeq
A term $\sim \int d^3 k$ at zero temperature is dropped, and certainly
vanishes when all such terms are summed together.
Expanding as in Eq.~(\ref{sum_dist_fncs}), with $Q = 2 \pi T q$,
\beq
\nAcal(Q) 
= \frac{3}{ \pi^3 } \; \sum_{j=1}^\infty \frac{1}{j^3} \; \sin(2 \pi q j)
=  q ( 1-q) ( 1 - 2 q) \; .
\label{scalar_k0_C}
\eeq
We recognize this as the {\it first} derivative of the potential in 
Eq.~(\ref{define_one_loop_potA}), 
\beq
\nAcal(Q) = \frac{d}{d q} {\cal V}(Q) \; .
\label{scalar_k0_D}
\eeq
Notice that this term manifestly vanishes when $Q = q = 0$.  It also
vanishes for $q = 1$, because this is a vacuum equivalent to $q = 0$,
and for $q= 1/2$, because this is an extremal point of the potential.

\subsection{Gluonic hard thermal loops: Landau damping}
\label{subsec_gluon_htls_landau}

We next turn to the nontrivial hard thermal loops, which have
discontinuities, as a function of the external momenta.  These arise
from energy denominators with Landau damping.

We start by deriving, briefly, how such hard thermal loops arise for
the gluon (or photon) self-energy when $Q=0$.  
Consider, as a prototype, the integral
\beq
\TInt \; \frac{k^i k^j}{K^2 (P-K)^2} \; .
\label{landau_A}
\eeq
For the quark contribution to the gluon self-energy, the factors
$k^i$ and $k^j$ arise from the quark propagators.  For the gluon 
contribution, these momenta arise from the momentum dependence of the
three gluon vertices.  There are, of course, other contributions,
with momenta $(k_0)^2$ and $k_0 k_i$, but these can be treated similarly.
We take the momenta in the numerator to be $\sim k$ because the loop
momenta for hard thermal loops are hard momenta, $k \sim T$, and dominate
over soft loop momenta, $\sim g T$.

This integral is done as for Eq.~(\ref{scalar_qk_self_energy}) in
Sec.~\ref{sec_quark_self_energy}.  There are two terms which
persist at zero temperature, ${\cal I}_1$ and ${\cal I}_4$, with
energy denominators $i p_0 \mp (E_k + E_{p - k})$.
If $k \sim T$ is hard, and $p \sim g T$ is soft, then these energy
denominators are hard, $\sim 2 k$.  Such terms are independent of
$p_0$ and $p$, so that $\mathcal{I}_{1}+\mathcal{I}_{4}\approx 2n(k)/k$,
which just produce a tadpole term as in Eq.~(\ref{tadpole_A}):
\beq
\frac{1}{8 \pi^2} \int \frac{d \Omega}{4 \pi} \; \hat{k}^i \hat{k}^j \;
  \int_0^\infty dk \; k^{2}
\;\frac{ 2\,n(k)}{k}
=\frac{\delta^{ij}T^{2}}{72} \;.
\label{landau_B}
\eeq

Instead, concentrate on the terms ${\cal I}_2$ and ${\cal I}_3$,
which arise from the denominators with Landau damping:
\beq
 \frac{1}{8 \pi^2}
 \int \frac{d \Omega}{4 \pi} \; \hat{k}^i \hat{k}^j \;
\;  \int_0^\infty dk \; k^2 \; \left( n(E_k) - n(E_{p - k})\right)
\left( \frac{1}{i p_0 - E_k + E_{p - k} }
- \frac{1}{i p_0 + E_k - E_{p - k} } \right) \; .
\label{landau_C}
\eeq
For the residues of the propagators we have taken $E_{p-k}\sim k$,
but this approximation cannot be made so cavalierly
in the rest of the expression.
Indeed, notice that without any cancellation, 
these terms are nominally
larger than we expect, $\sim \int dk \; k^2 \; n(k) \sim T^3$,
and not $\sim T^2$.  However, 
what enters into Eq.~(\ref{landau_C}) is only 
the {\it difference} of the statistical distribution functions.
For $k\sim T$ and $p \sim gT$,
\beq
n(E_{p-k}) -  n(E_k) \approx
\frac{\vec{p} \cdot \hat{k}}{T} \; n(k) \left(1 + n(k) \right) \; .
\label{landau_D}
\eeq
Because of this cancellation in the statistical distribution functions,
the diagram is not $\sim T^3$, but only $\sim T^2$.  It is a product
of an integral over $k$,
\beq
\frac{1}{4 \pi^2 T} \int_0^\infty dk \; k^2
\; n(k) \left( 1 + n(k) \right) \; , 
\label{landau_E}
\eeq
and an angular integral,
\beq
- \frac{1}{2}\int \frac{d \Omega}{4 \pi} \; \hat{k}^i \hat{k}^j \;
(\hat{p} \cdot \hat{k}) \; 
\left( \frac{1}{i p_0/p - \hat{p} \cdot \hat{k}} 
- \frac{1}{i p_0/p + \hat{p} \cdot \hat{k}}  \right)
\; .
\label{landau_F}
\eeq
This is the usual hard thermal loop.  The integral over $k$
generates the thermal mass for the gluon.  This is multiplied times
an angular integral, which generates a dimensionless function of $ip_0/p$.
This function has discontinuities on the light
cone from the Landau damping of massless particles.

We write the angular integral in Eq.~(\ref{landau_F}) as we do to
emphasize its behavior as a function of 
\beq
x = \cos \theta = \hat{p} \cdot \hat{k} \; .
\label{landau_G}
\eeq
The integral over $x$ is from $-1$ to $+1$, so a nonzero result must be
even in $x$.  In Eq.~(\ref{landau_F}), this happens because the difference
of statistical distribution functions, and the difference of energy
denominators, are each odd in $x$, so the product is even.  

Now consider the analogous integral when $Q \neq 0$.  As a typical
example, consider
\beq
{\cal J}^{i j}(P^{12},Q_1,Q_2) =\frac{1}{4} \TInt \; 
\frac{(2k^i -p^i) \;(2 k^j-p^j) }{(K + Q_1)^2 (P - K + Q_2)^2} \; \; ; \;\;
P^{1 2} = P + Q_1 + Q_2 \; .
\label{landauQ_A}
\eeq
Here $k_0$ and $p_0$ are both taken to be bosonic momenta, while $Q_1$ and
$Q_2$ are arbitrary background fields.  The numerator is chosen to $(2k^i -p^i) \;(2 k^j-p^j)$, which
differs terms proportional to $p^{i}$ and $p^{j}$ from Eq.~(\ref{landau_A}). 
We need this term to keep a symmetry, $Q_1\leftrightarrow Q_{2}$.
This symmetry can be checked by shifting of integration variables, $K\to P-K$,
since both $k_0$ and $p_0$ are, by assumption, bosonic momenta.
In the numerator we take
$\sim k^i k^j$ as a term which generates the largest terms for hard $k$;
terms $\sim k^i p^j$ are down by $\sim g$ to this term, 
those $\sim p^i p^j$ by $\sim g^2$.  We keep the numerator of order $p$
because the existence of the background field makes the new leading order, which we will see in the following, of order $g^{2}T^{3}/p$.  If one neglects the terms proportional to $p^{i}$ and $p^{j}$ in the numerator of Eq.~(\ref{landauQ_A}), one find a new term which is odd under $Q_{1}\leftrightarrow Q_{2}$.
To see this symmetry in another way, note that the relevant
external momentum is $P^{1 2} = P + Q_1 + Q_2$.  Then the hard thermal
loops in Eq.~(\ref{landauQ_A}) can be rewritten as
\beq
\frac{1}{4}\TInt \; 
\frac{(2k^i -p^{i})\; (2k^j-p^{j})}{(K^1)^2 (P^{1 2} - K^1)^2} =
\frac{1}{4}\TInt \; 
\frac{(2k^i -p^{i})\; (2k^j-p^{j})}{(K^2)^2 (P^{1 2} - K^2)^2} \; ,
\label{landauQ_A3}
\eeq
where $K^1 = K + Q_1$ and $K^2 = K + Q_2$. This is not a $Q$-dependent
shift of momenta, which would be invalid for arbitrary $Q$.  Instead,
$K^1 \rightarrow P^{1 2} - K^2$ is
just a shift of purely bosonic momenta,
$K+Q_1 \rightarrow P - K + Q_1$.

In Eq.~(\ref{landauQ_A}) there are terms 
which persist at zero temperature, ${\cal I}_1$ and ${\cal I}_4$,
with energy denominators $i p_0^{12} \mp (E_k + E_{p - k})$.
These are really no different than for $Q=0$, Eq.~(\ref{landau_B}); the dependence on
the external momenta drops out, and one is left with integrals
\beq
\begin{split}
\sim &\; \int \frac{d^3 k}{(2 \pi)^3}
\; \frac{k^i k^j}{4k^2} \; 
\frac{1}{2k}\left( n(k - i Q_1) + n(k + i Q_{1})+ n(k - i Q_2) + n(k + i Q_{2}) \right) \\
 =& \frac{T^2\delta^{ij}}{144} \;(\Acal(Q_{1})+\Acal(Q_{2}))
 \; ,
 \end{split}
\label{landau_A1}
\eeq
where we have dropped $p^{i}$ on the numerator, which is higher order of hard thermal loop approximation. 
This is just like the integral of Eq.~(\ref{tadpole_B}).
The angular integral is trivial, $\sim \delta^{i j}$,
and the $Q$-dependence is only through terms $\sim T^2 \Acal(Q)$.
(As noted before, tadpole integrals such as Eq.~(\ref{scalar_k0_A})
arise only for the one point gluon function, and not in the gluon
propagator.  This is clear just on dimensional grounds, as that diagram
has dimensions of $({\rm mass})^3$.)

We then turn to the terms with energy denominators which correspond
to Landau damping, ${\cal I}_2$ and ${\cal I}_3$.  This part of the
integral is
\beq
{\cal J}^{i j}(P^{12}, Q_1, Q_2) = 
\frac{1}{8 \pi^2} \; \int^\infty_0 dk \; \frac{k^4}{E_k E_{p - k}} \;
 \int \frac{d \Omega}{4 \pi} \; \left(\hat{k}^i -\frac{p^{i}}{2k}\right)\left(\hat{k}^j -\frac{p^{j}}{2k}\right)
\left( {\cal I}_2 + {\cal I}_3 \right) \; .
\label{landauQ_B1}
\eeq
The statistical distribution functions, and energy denominators, which
represent Landau damping are the generalization of that for
$Q=0$, Eq.~(\ref{landau_C}), to $Q \neq 0$.
These are just modifications of
Eqs.~(\ref{first_integral_F}) and (\ref{first_integral_G}), replacing
the $Q$'s there by $Q_1$ and $Q_2$:
\beq
{\cal I}_2 + {\cal I}_3  =
\frac{n(E_k - i Q_1) - n(E_{p - k} + i Q_2) }
{ip_0^{1 2} - E_k + E_{p - k}}
+\; \frac{n(E_{p-k} - i Q_2) - n(E_{k} + i Q_1) }
{ip_0^{1 2} + E_k - E_{p - k}} \; ;
\label{landauQ_C}
\eeq
$p_0^{1 2} = p_0 + Q_1 + Q_2$.
Now we symmetrize each term with respect to then interchange of $Q_1$
and $Q_2$:
\beq
{\cal I}_2 \; = \; \frac{1}{2} \;
\frac{n(E_k - i Q_1) - n(E_{p - k} + i Q_1)
+ n(E_k - i Q_2) - n(E_{p - k} + i Q_2) }
{ip_0^{1 2} - E_k + E_{p - k}} \; ,
\label{landauQ_DA}
\eeq
\beq
{\cal I}_3 \; = \; \frac{1}{2} \;
\frac{ n(E_{p-k} - i Q_1)  - n(E_{k} + i Q_1) 
+ n(E_{p-k} - i Q_2) - n(E_{k} + i Q_2) }
{ip_0^{1 2} + E_k - E_{p - k}} \;.
\label{landauQ_DB}
\eeq

After symmetrization it is then easy to pick out both the leading,
and next to leading terms from
(\ref{landauQ_DA}) and (\ref{landauQ_DB}).  

The leading term is easy.  In every term, we
approximate $E_{p - k} \approx E_k = k$, and neglect $p^{i}$ and $p^{j}$ in
the numerator, so we find that
${\cal J}^{i j}(P, Q_1, Q_2)$ factorizes into a product of an integral
over $\int dk$, and an angular integral.  The former is
$$
\frac{1}{16 \pi^2}
\int^\infty_0 dk \; k^2
\left( n(k - i Q_1) - n(k + i Q_1) + n(k - i Q_2) - n(k + i Q_2)
 \right)
$$
\beq
= \frac{i \pi T^3}{12} \; \left( \nAcal(Q_1) + \nAcal(Q_2) \right) 
\; ,
\label{landauQ_E}
\eeq
using the function $\nAcal$ of Eq.~(\ref{scalar_k0_C}).  
In all,
\beq
{\cal J}^{i j}(P^{12},Q_1,Q_2)
= 
\frac{\pi T^3}{12p} \; \left( \nAcal(Q_1) + \nAcal(Q_2) \right) 
\int \frac{d \Omega}{4 \pi} \; \hat{k}^i \hat{k}^j 
\left( \frac{1}{i p_0^{1 2}/p - \hat{p} \cdot \hat{k}} 
+ \frac{1}{i p_0^{1 2}/p + \hat{p} \cdot \hat{k}}  \right) \; .
\label{landauQ_F}
\eeq
In the angular integral, $\hat{k}^i \hat{k}^j$ produce terms $\sim 1$
and $x^2$, where $x = \hat{p}\cdot \hat{k}$, Eq.~(\ref{landau_G}).
The angular integral is manifestly even in $x$, 
and so does not vanish.
This expression can be rewritten in a form similar
to that of Eq.~(\ref{htl_quark_A}),
\beq
{\cal J}^{i j}(P^{12},Q_1,Q_2)
\HTLapprox
\frac{\pi T^3}{6} \; \left( \nAcal(Q_1) + \nAcal(Q_2) \right) 
\delta \varGamma^{i j}(P^{1 2})  + \ldots
\; .
\label{landauQ_FA}
\eeq
We introduce the function
\beq
\delta \varGamma^{\mu \nu}(P) 
= \; \int \frac{d \Omega}{4 \pi} \;
\left( \frac{ \hat{K}^\mu \hat{K}^\nu }
{P \cdot \hat{K}}  \right) \; ;
\label{landauQ_FB}
\eeq
$\hat{K} = (i, \hat{k})$, with $d\Omega$ the integral over $\hat{k}$.
The hard thermal loop is $g^2$ times ${\cal J}^{i j}$.  Since
$\delta \varGamma \sim 1/P$, this is
$\sim g^2 T^3/p$ times a dimensionless function of $ip_0^{1 2}/p$.
This is rather different from the hard thermal loops for $Q=0$,
which are $\sim g^2 T^2$ times a dimensionless function
of $ip_0/p$.

The origin for this difference is natural, when one considers the
propagator in a background field, Eq.~(\ref{field_strength}).
There are the ordinary terms, $\sim (D^\text{cl}_\mu)^2$, which becomes
$(P^{a b}_\mu)^2$ in momentum space.  In addition, however,
there is also a term $\sim 2 i g [G^\text{cl}_{\mu \nu}, .]$.  For
a $\ZN$ interface, 
\beq
g \; G^\text{cl}_{0 z} \sim g \; \partial_z \left( \frac{T q(z)}{g} \right)
\sim T \; \partial_z q(z) \sim g \; T^2 \; ,
\label{landauQ_G}
\eeq
where we use the fact that the typical spatial momenta for a $\ZN$
interface is small, $\sim g T$.  Thus this term is {\it larger}
by $\sim 1/g$ than the ordinary terms, 
$\sim P^2 \sim g^2 T^2$.  

The hard thermal loop in Eq.~(\ref{landauQ_F}) is $\sim g^2 T^3/p$,
which for soft $p$ is $\sim g T^2$.  Thus the new hard thermal loop
can be viewed as a modification of the term in the background
field propagator for a gluon.  Like any other hard thermal loop,
it is as large as the term at tree level for soft external momenta.
It is just that in a background field, this term is larger than
expected.  

The term $\sim 2 i g [G^\text{cl}_{\mu \nu}, .]$ is special to a non-Abelian
gauge field, since it involves the commutator in group space. It also
has no analogy for a fermion field, either Abelian or non-Abelian.  
This explains why it did not appear in previous examples.

Having such a term is special to computing for a background field with
a nonzero color field; for a $\ZN$ interface, it is a nonzero color
electric field.  Thus for the semi-QGP, one does not expect a nonzero
color field in vacuum, and such terms should not appear.  We admit
that at present, we do not have a fully self-consistent theory of the 
semi-QGP, which would allow us to demonstrate this.

We turn to subleading terms in the gluon self-energy, $\sim T^2$.
In this, it helps greatly to recognize that the integral must be even
in $x = \cos \theta$, Eq.~(\ref{landau_G}).  The measure
is even in $x$, as are terms $\sim k^i k^j$, which
produce contributions $\sim 1$ or $\sim x^2$.

There are several ways that corrections $\sim T^2$ can arise.
The first is a numerator proportional to $p^{i}$ and $p^{j}$:
\beq
\begin{split}
&-\frac{1}{32 \pi^2} \; \int^\infty_0 dk \; k \;
\left( n(k - i Q_1) - n(k + i Q_1) + n(k - i Q_2) - n(k + i Q_2)\right) \\
&\qquad\times  \int \frac{d \Omega}{4 \pi} \; \left(\hat{k}^i p^{j}+p^{i} \hat{k}^j\right)
\left( \frac{1}{i p_0^{1 2}/p - \hat{p} \cdot \hat{k}} 
+ \frac{1}{i p_0^{1 2}/p + \hat{p} \cdot \hat{k}}  \right) \; .
\end{split}
\label{landauQ_B2}
\eeq
This vanishes, because the integrand is odd in $\hat{k}$.
The second is expanding $1/E_{p - k}$ which arises in the measure of the
integral, as the residue of the propagator,
\beq
\frac{1}{E_{p-k}} \sim \frac{1}{k} + \frac{\vec{p} \cdot \hat{k}}{k^2 } 
+ \ldots \; .
\label{landauQ_H}
\eeq
Again, since $\vec{p} \cdot \hat{k} = p x$, this is odd in $x$, and so vanishes.

The third is by expanding $E_{p-k}$ in the energy denominators:
\beq
\frac{1}{i p_0^{1 2} \pm (E_k - E_{p-k})}
\approx 
\frac{1}{i p_0^{1 2} \pm \vec{p} \cdot \hat{k}} \mp
\frac{\vec{p}^{\; 2} - \left( \vec{p} \cdot \hat{k}\right)^2}
{2 k\left( i p_0^{1 2} \pm \vec{p} \cdot \hat{k}\right)^2}
+ \ldots
\label{landauQ_I}
\eeq
The numerator of the second term on the right hand side is $p^2 (1 - x^2)$,
which is even in $x$.  Because of the $\mp$ sign in front of the second
term, though, this is in all odd in $x$, and so vanishes.

Thus, the {\it only} way that corrections $\sim T^2$ arise is by
expanding $E_{p - k}$ in the statistical distribution functions
\beq
n(E_{p - k} - i Q)
\approx n(k - i Q) + \frac{\vec{p} \cdot \hat{k}}{T}\;
n(k - i Q) \left( 1 + n(k - i Q) \right) \; + \ldots
\label{landauQ_J}
\eeq
This is exactly the same sort of expression as at $Q=0$.  Now this
term is odd in $x$, but note that 
$- n(E_{p -k} + i Q_{1,2})$ enters in ${\cal I}_2$, 
Eq.~(\ref{landauQ_DA}), and 
$+ n(E_{p -k} - i Q_{1,2})$ enters in ${\cal I}_3$, Eq.
(\ref{landauQ_DB}).  Thus the result is even in $x$, as it must
be if not to vanish.  Explicitly, the terms $\sim T^2$ in 
${\cal J}^{i j}$ are a product of an integral over $k$,
\beq
\frac{1}{16 \pi^2 T}
\int_0^\infty dk \; k^2 \left(
n(k- i Q_1) (1 + n(k - i Q_1) )
+ n(k + i Q_1) \left(1 + n(k + i Q_1) \right) + (Q_1 \leftrightarrow Q_2) \right) \;,
\label{landauQ_K}
\eeq
and an integral over the angular variables,
\beq
\int \frac{d \Omega}{4 \pi} \;
\left( \frac{ \hat{k}^i \hat{k}^j \; \vec{p} \cdot \hat{k}}
{P \cdot \hat{K}}  \right) \; .
\label{landauQ_L}
\eeq

The angular integral in Eq.~(\ref{landauQ_L}) is identical to that
for ordinary hard thermal loops.  The momentum integral is also
a minor modification.  Consider
\beq
\frac{1}{T}\;
\int_0^\infty dk \; k^2 
\left( n(k - i Q) (1 + n(k - i Q) )
+ n(k + i Q) (1 + n(k + i Q) \right)
\; ,
\label{landauQ_M}
\eeq
which we can rewrite as 
\beq
- i \; \frac{\partial}{\partial{Q}}
\int_0^\infty dk \; k^2 
\left( n(k - i Q) - n(k + i Q) \right)
\; .
\label{landauQ_N}
\eeq
This integral arose previously in Eq.~(\ref{scalar_k0_C}), and
involves the function $\nAcal(Q)$, which is
the first derivative of the potential
${\cal V}(Q)$, Eq.~(\ref{define_one_loop_potA}).  
Since in Eq.~(\ref{landauQ_N})
we take a derivative of this function with respect to $Q$, however,
the momentum integral in Eq.~(\ref{landauQ_M}) involves not
the first derivative of ${\cal V}(Q)$, but the second, through the
function $\Acal(Q)$, Eq.~(\ref{define_one_loop_potB}):
\beq
\frac{2 \pi^2 T^2}{3} \; \Acal(Q) \; .
\label{landauQ_O}
\eeq
In summary, the terms $\sim T^3$ in ${\cal J}^{i j}(P,Q_1,Q_2)$ are
those of Eq.~(\ref{landauQ_FA}); the terms $\sim T^2$ are 
\beq
\frac{T^2}{24}\;
\left( \Acal(Q_1) + \Acal(Q_2) \right)
\left( \frac{\delta^{i j}}{6}
+ \int \frac{d \Omega}{4 \pi} \;
\frac{ \hat{k}^i \hat{k}^j \; \vec{p} \cdot \hat{k}}
{P^{1 2} \cdot \hat{K}}  \right) \; .
\label{landauQ_P}
\eeq
For completeness, we have added the tadpole terms, $\sim \delta^{i j}/6$.
This expression can be rewritten as
\beq
\frac{T^2}{24}\;
\left( \Acal(Q_1) + \Acal(Q_2) \right)
\left(\frac{\delta^{ij}}{2}-
i p_0^{1 2}\int \frac{d \Omega}{4 \pi} \;
\frac{ \hat{k}^i \hat{k}^j }
{P^{1 2} \cdot \hat{K}}  \right) \; .
\label{landauQ_Q}
\eeq

It is direct to show that these are the only terms $\sim T^3$ or $\sim T^2$.
The one concern is terms where the numerator is $\sim k^i p^j$ or
$\sim p^i k^j$: while the single power of $p^i$ brings in a suppression
by a soft momenta, $\sim p/T$,
this times a term $\sim T^3/p$ could produce a result
$\sim T^2$.  However, a single power of $k^i$ or $k^j$ is manifestly
odd in the angular variable $x$, Eq.~(\ref{landau_G}).  In contrast,
the terms which produce contributions $\sim T^3$, Eq.~(\ref{landau_F}),
are even in $x$.  Thus these possible terms $\sim T^2$ vanish when
integrated over $x$. Therefore, in the hard thermal loop approximation, terms 
proportional to $p^i$ and $p^j$ in Eq.~(\ref{landauQ_A})  may simply be dropped,
\beq
{\cal J}^{i j}(P^{12},Q_1,Q_2) \HTLapprox  \TInt \; 
\frac{k^i \;k^j }{(K + Q_1)^2 (P - K + Q_2)^2} \; .
\eeq

For the gluon self-energy, the general integral required is
\beq
\begin{split}
{\tilJ}^{\mu \nu}(P^{1 2},Q_1,Q_2) =& \frac{1}{4} \, \TInt \; 
\left( - \; \delta^{\mu \nu} \;
\left( \frac{1}{(K + Q_1)^2} 
  + \frac{1}{(K + Q_2)^2 } \; \right)\right.  
\\& \left.
\qquad\qquad\quad+ \; \frac{(2 K - P + Q_1 - Q_2)^{\mu} \; 
(2 K - P + Q_1 - Q_2)^\nu}{(K + Q_1)^2 (P - K + Q_2)^2} 
 \right)\; ,
\end{split}
\label{landauQ_PP}
\eeq
where $Q^\mu = u^\mu Q$, $u^\mu = 1$ if $\mu = 0$, and zero otherwise.
The first term on the left hand side, $\sim \delta^{\mu \nu}$, is added
in anticipation of the integrals which arise for hard thermal loops,
and is obviously independent of the external momentum, $P$.

Momentum dependence arises from the second term on the left hand side.
Its numerator involves
$2K - P + Q_1 - Q_2 = 2 K^1 - P^{1 2}$;
under a shift of the loop momentum, $K \rightarrow P-K$, this becomes
$- (2 K^2 - P^{1 2} )$.  Comparing with
Eq.~(\ref{landauQ_A3}) shows that Eq.~(\ref{landauQ_PP}) is symmetric
under interchange of $Q_1$ and $Q_2$.  

When $Q = 0$, the computation of hard thermal loops is simplified
by keeping only powers of the hard loop momentum
$K^\mu$ in the numerator, and dropping 
those of the soft external momenta, $P^\mu$ \cite{htlA,htlB,lebellac}.  
When $Q \neq 0$, the analogous approximation is to keep powers
of $K^1$ or $K^2$, and drop powers of $P^{1 2}$, as we saw in ${\cal J}^{ij}$.
It is essential to remember that while $Q_1$ and $Q_2$ are each
hard, that the real energy $P^{1 2}$ is soft, Sec.~\ref{sec_analytic_cont}.
Thus we can write the second term on the left hand side of Eq.
(\ref{landauQ_PP}) as
\beq
\TInt \; 
\frac{(K^1)^{\mu} \; (K^1)^\nu}{(K^1)^2 (P^{1 2} - K^1)^2} 
\HTLapprox
\TInt \; 
\frac{(K^2)^{\mu} \; (K^2)^\nu}{(K^2)^2 (P^{1 2} - K^2)^2} \; .
\label{landauQ_P3}
\eeq
This simplification will help greatly in computing the gluon self-energy
in Secs.~\ref{quarkcontgluonselfenergy} and
\ref{gluoncontgluonselfenergy}.

Computations similar to those above show that this integral equals
\beq
{\tilJ}^{\mu \nu}(P^{1 2},Q_1,Q_2)
\HTLapprox \frac{i \pi T^3}{6} \; \left( \nAcal(Q_1) + \nAcal(Q_2) \right) 
\delta \varGamma^{\mu \nu}(P^{1 2})
+ \frac{T^2}{24}\;
\left( \Acal(Q_1) + \Acal(Q_2) \right)
\delta \varPi^{\mu \nu}(P^{1 2}) \; ,
\label{landauQ_R}
\eeq
where the momentum dependence enters through the functions
$\delta \varGamma^{\mu \nu}(P^{1 2})$, Eq.~(\ref{landauQ_FB}), and 
\beq
\delta \varPi^{\mu \nu}(P) = 
\; \left( - u^\mu u^\nu - i p_0\int \frac{d \Omega}{4 \pi} \;
\frac{ \hat{K}^\mu \hat{K}^\nu }
{P \cdot \hat{K}}  \right) \; .
\label{landauQ_T}
\eeq
Again, $\hat{K} = (i, \hat{k})$, so $\hat{K}^2 = 0$, and the angular integral
is over all directions of $\hat{k}$.  

Previously we computed $\delta \varGamma^{i j}$ and $\delta \varPi^{i j}$.
Given this result, it is immediate to show that
$\delta \varGamma^{i 0}$ and $\delta \varGamma^{ 0 0}$ are correct.
The only effort necessary is to establish the correctness
of $\delta \varPi^{i 0}$ and $\delta \varPi^{ 0 0}$.
However, the derivation of $\delta \varPi^{i j}$ 
above shows that the terms $\sim T^2$ arise in precisely the same way when
$Q \neq 0$ as for $Q = 0$, entirely from the change in the statistical
distribution functions, Eq.~(\ref{landauQ_J}).  Thus we can be
certain that the only change is that of the thermal gluon mass in 
a background field.  That is, the momentum dependence, which enters
through $\delta \varPi^{\mu \nu}(P)$, is unchanged.  For this,
the previous computation suffices.

The function $\delta \varPi^{\mu \nu}(P)$ is the same function as arises
in the hard thermal loops when $Q = 0$.  The function 
$\delta \varGamma^{\mu \nu}(P)$ enters only when $Q \neq 0$, since 
it is multiplied by $\nAcal$, and $\nAcal(0) = 0$.  
From Eqs.~(\ref{landauQ_FB}) and (\ref{landauQ_T}), the two functions
are simply related to one another,
\beq
\delta \varGamma^{\mu \nu}(P) = \frac{-1}{ip_0}
\left( \delta \varPi^{\mu \nu}(P) + u^\mu u^\nu \right) \; .
\label{landauQ_T1}
\eeq
The function $\delta \varPi^{\mu \nu}(P)$ is transverse,
\beq
P^\mu \; \delta\varPi^{\mu \nu}(P) = 0 \; .
\label{landauQ_T2}
\eeq
Thus the new hard thermal loop is not,
\beq
P^\mu \; \delta\varGamma^{\mu \nu}(P) =  i u^{\nu}\; .
\label{landauQ_T3}
\eeq
As discussed following Eq.~(\ref{landauQ_G}), the new hard thermal loop
is the modification of the propagator in a background field.  Thus it
is not necessary for the term to be transverse.  We shall also see this
in the next section, where we compute the one point function for a gluon.

\subsection{One point gluon function}
\label{sec_one_point_gluon}

With these results in hand, the diagrams can be computed directly.  In this
subsection we begin with the one point function for a gluon.  
The integral which enters is that of Eq.~(\ref{scalar_k0_A}), and
generates the function $\nAcal$, Eq.~(\ref{scalar_k0_C}).
As can be seen from Eq.~(\ref{scalar_k0_B}), this function vanishes
when $Q=0$, as then the integral is odd in $k_0$.

Consider a gluon loop tied onto a gluon line.
By an argument similar to that which lead to
Eqs.~(\ref{insert_gluon_quark_line}) and 
(\ref{insert_gluon_quark_line_simplify}), 
if the gluon propagator in the loop is
${\cal P}^{cd, ef}/(K^{c d})^2$, Eq.~(\ref{gluon_prop}),
we can replace this by $\delta^{c f}\delta^{d e}/(K^{c d})^2$.
Since the external gluon has zero momentum, the three gluon vertex
involves one momentum, which we choose as 
$K^{c d}$; from Eq.~(\ref{three_gluon}), the three gluon vertex is
\beq
-i g f^{(ab, cd, ef)}
\left( -K_\lambda^{cd} \; \delta_{\mu \nu}
+ 2 K_\mu^{c d} \; \delta_{\nu \lambda}
- K^{c d}_\nu \; \delta_{\lambda \mu} \right) \; .
\label{tadpole_three_gluon}
\eeq
This is transverse in $K_\nu^{cd}$ and $K_\lambda^{cd}$,
so the gauge dependent term in the gluon propagator, 
$\sim( \xi - 1)K_\nu^{cd}K_\lambda^{cd}$,
Eq.~(\ref{gluon_prop}), drops out.
The term in the gluon propagator $\sim \delta^{\mu \nu}$ acting
upon Eq.~(\ref{tadpole_three_gluon}) gives $2(d-1)$ in $d=4$
spacetime dimensions.  Including an overall $1/2$ for the gluon loop,
\beq
-\, \frac{6}{2}\, i\, g \;
f^{ab,cd,dc}
\; \TInt \; \frac{(K^{cd})^\mu}{(K^{cd})^2}
= -  \, u^\mu \, i \, g \, \pi \, T^3 \; \sum_{c,d=1}^\Nc
\; f^{ab, cd, dc} \; T^3 \; \nAcal(Q^c-Q^d) \; .
\label{gluon_tadpole_A}
\eeq
Here $u^\mu = (1,\vec{0}\,)$.
The contribution from a ghost loop is similar, with a coefficient of
$-1$ instead of $+3$ in Eq.~(\ref{gluon_tadpole_A}); thus the sum is
$+2$, which reflects for the two transverse degrees of freedom of a 
gluon.

The contribution of $\Nf$ flavors of massless quarks is
\beq
(-) i \, g \; \Nf \;
\left(t^{ab}\right)_{c c}\,\TInt\; {\rm tr} \;
\frac{\gamma^{\mu} }{-i \otherSlash \cKtl }
=  \frac{4\pi}{3} \, u^\mu \, g \; \Nf\; T^3 \,  \left(t^{ab}\right)_{cc} \,
\nAcal (\cQtl ) \; .
\label{quark_tadpole_B}
\eeq

Using Eqs.~(\ref{generator_projector}) and (\ref{three_gluon_vertex}),
the sum of the gluon, ghost, and quark contributions is
\beq
\langle J^{a b}_\mu \rangle \HTLapprox
- \, u^\mu\; \delta^{a b} \; \frac{4\,\pi \, g \, T^3}{3 \, \sqrt{2}}
\left(
 \sum_{c = 1}^\Nc \; \left[
\nAcal\left(Q^{ac} \right)
+ \frac{\Nf}{\Nc} \; \nAcal(\cQtl)
\right]
 - \Nf \;\nAcal (\aQtl)
\right) \; .
\label{total_tadpole}
\eeq
In obtaining this, we have used the fact that $\nAcal$ is odd
in $Q \rightarrow - Q$.  This is clear from its definition
in Eq.~(\ref{scalar_k0_B}); its form in Eq.~(\ref{scalar_k0_C}) 
is only valid for $0 < q < 1$.

For a $\ZN$ interface, this current is part of the equation of motion
for the gluon, corrected to one-loop order.  There it is natural:
the interface arises from a balance of the Lagrangian at tree level,
and a potential for the $Q$'s at one-loop order.  Equation~(\ref{total_tadpole})
is precisely the derivative of the one-loop potential.

For the semi-QGP, a term
must be added to the Lagrangian to cancel the contribution of this
current.  This is natural; if such a term is not added, the minimum
would be at $Q = 0$, or equivalent points; i.e., the usual $\ZN$ minima.

\subsection{Quark contribution to the gluon self-energy}
\label{quarkcontgluonselfenergy}

With the previous examples in hand, the computation of the gluon self-energy 
is mainly a matter of putting things together.  Even so,
because the $Q$'s are nontrivial, what is of interest is to see how
new color structures arise at one-loop order from quantum corrections
to the propagators at tree level.  This is unlike the case of the quark
self-energy, Eq.~(\ref{htl_quark_A}), where the only change from $Q = 0$
was the value of the thermal quark mass.

\begin{figure}
\includegraphics[width=0.75\textwidth]{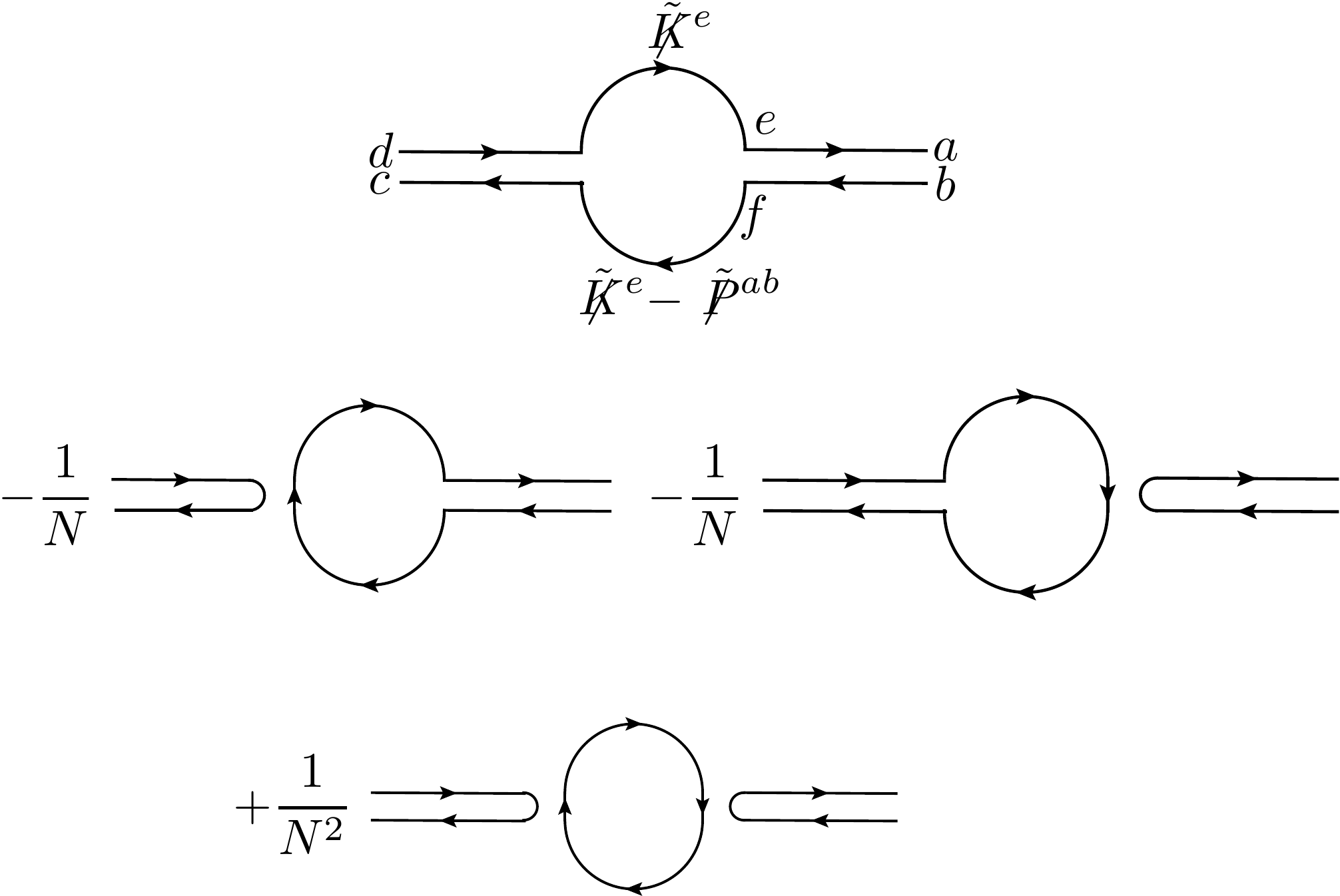}
\caption{One-loop diagram for the quark loop in the gluon self-energy.}
\label{fig_quark_loop_gluon_self_energy}
\end{figure}

We start with the contribution of the quark loop.  For $\Nf$ flavors,
this is
\beq
-(\varPi^{ab,cd}_{\mu \nu})_{\rm qk}(P^{a b})
= (-) \; (ig)^2 \, \Nf \; \TInt
\; {\rm tr}\;
\;\gamma^\mu \; (t^{a b})_{e f} \; 
\frac{1}{-i(\otherSlash \eKtl)} 
\; \gamma^\nu \; (t^{c d})_{f e}
\frac{1}{-i(\otherSlash \eKtl - \otherSlash P^{a b})}  \; .
\label{gluon_self_quark1}
\eeq
Keeping only the terms $\sim \otherSlash \eKtl$ in the numerators, the hard
thermal loop in this contribution is
\beq
(\varPi^{ab,cd}_{\mu \nu})_{\rm qk}(P^{a b})
\HTLapprox \
 \, 8 \, g^2 \, \Nf \; 
\sum_{e,f=1}^{\Nc} \; (t^{a b})_{e f} \; (t^{c d})_{f e} \;
{\tilJ}^{\mu \nu}(P^{a b},\eQtl,Q^{ab}- \eQtl) \; .
\label{gluon_self_quark2}
\eeq
Because $\eQtl$ appears in the function $\tilJ^{\mu\nu}$, 
what enters is not simply the trace of two projection operators,
Eq.~(\ref{trace_gen}).  The sum over the color index $e$ and $f$
is easy, and gives 
\beq
\begin{split}
(\varPi^{ab,cd}_{\mu \nu})_{\rm qk}(P^{a b})
\HTLapprox \ &
 \, 4 \, g^2 \, \Nf \; 
\Biggl[
\delta^{a d} \delta^{b c} \; {\tilJ}^{\mu \nu}(P^{a b},\aQtl,-\bQtl)  \\
&\qquad\quad-\frac{1}{\Nc} \; \delta^{a b} \delta^{c d} \Bigl(
{\tilJ}^{\mu \nu}(P, \aQtl,-\aQtl)
+{\tilJ}^{\mu \nu}(P, \cQtl,-\cQtl)\\
&\qquad\qquad\qquad\qquad- \; \frac{1}{\Nc} 
\; \sum_{e=1}^{\Nc} {\tilJ}^{\mu \nu}(P, \eQtl,-\eQtl) \Bigr) \Biggl] \; .
\label{gluon_self_quark3}
\end{split}
\eeq
This is illustrated in Fig.~\ref{fig_quark_loop_gluon_self_energy}.
There is the usual planar diagram, plus three contributions from
diagrams in which one or both of the gluon indices are traced.  
Notice that as usual, the color structure is far more clear from
the diagram, than from the detailed expression in
Eq.~(\ref{gluon_self_quark3}).

The last three terms, $\sim \delta^{a b} \delta^{c d}$, are like
the photon propagator for QED, where the fermions 
propagate in a background $Q$ field.
In all, the quark contribution is traceless,
\beq
\sum_{a=1}^{\Nc} \; (\varPi^{a a,cd}_{\mu \nu})_{\rm qk}(P)
= \sum_{c=1}^{\Nc} (\varPi^{ab,cc}_{\mu \nu})_{\rm qk}(P) = 0 \; .
\label{gluon_self_quark4}
\eeq
which is necessary for self-consistency.
Note that the quark contribution is not simply proportional to a 
projector operator, $\sim {\cal P}^{a b,cd}$.  Instead, because
the $Q$'s can be unequal, at one-loop order the color
structure is more complicated, Eq.~(\ref{gluon_self_quark3}).

\subsection{Ghost and gluon contributions to the gluon self-energy}
\label{gluoncontgluonselfenergy}

To compute the contributions of gluons and ghosts to the hard
thermal loop in the gluon self-energy, we first compute in background
Feynman gauge, $\xi = 1$ in Eq.~(\ref{gluon_prop}).  We then
show that the results are independent of $\xi$, in complete analogy
to when $Q = 0$.

The simplest contribution is the tadpole diagram, which is independent
of the external momentum.  This involves the four gluon vertex of 
Eq.~(\ref{four_gluon}).  For the gluon in the loop, we can use the
simplification of Eq.~(\ref{insert_gluon_quark_line_simplify}) to
take the gluon propagator as a Kronecker delta.  The result is
\beq
- \; \frac{g^2}{2} \; 6 \; f^{(ab,ef,gh)} \; f^{(cd,fe,hg)}
\; \TInt \; \delta^{\mu \nu} \; \frac{1}{(K^{fe})^2} \; .
\label{gluon_selfA}
\eeq
The coefficient $6 \rightarrow 2(d-1)$ in $d$ spacetime dimensions; the
$1/2$ is for a bosonic loop.  

The gluon and ghost loops are, generally, involved.  However, we can use
the simplification of Eqs.~(\ref{landauQ_P3}): write the momenta in terms
of the external momentum, $P^{a b}$, and a loop momentum, which we can
define as $K^{f e}$.  Then although the $Q$'s are hard to begin with,
we can consistently treat $P^{a b}$ as soft, and $K^{f e}$ as hard.
This allows us to drop powers of $P^{a b}$ uniformly.  (This is only valid
in Feynman and Coulomb gauges \cite{htlA,htlB,lebellac}).  

With this approximation it is then easy to read off the hard thermal loops
in the gluon self-energy.  The ghost loop is given by taking $K^{f e}$
at each vertex, and so is
\beq
(-) \;g^2 \; f^{(ab,ef,gh)} \; f^{(cd,fe,hg)} \; \TInt
\; \frac{(K^{f e})^\mu (K^{f e})^\nu}{(K^{f e})^2 (P^{a b} - K^{f e})^2} \; .
\label{gluon_selfB}
\eeq
For the contribution to the gluon self-energy from the diagram with two 
three gluon vertices, in each vertex we can neglect the external momentum, 
as in Eq.~(\ref{tadpole_three_gluon}). This gives
\beq
\;\frac{g^2}{2} \; f^{(ab,ef,gh)} \; f^{(cd,fe,hg)} \; \TInt
\; \frac{ 10 (K^{f e})^\mu (K^{f e})^\nu + 2(K^{f e})^2 \delta^{\mu\nu}}
{(K^{f e})^2 (P^{a b} - K^{f e})^2} \; .
\label{gluon_selfC}
\eeq
The coefficient $10 \rightarrow 2(2d-3)$ in $d$ spacetime dimensions.

In all, the sum of Eqs.~(\ref{gluon_selfA}), (\ref{gluon_selfB}),
and (\ref{gluon_selfC}) is
\beq
-(\varPi^{ab,cd}_{\mu \nu})_{\rm gl}(P^{a b}) \HTLapprox
\; 4 \; g^2 \; f^{(ab,ef,gh)} \; f^{(cd,fe,hg)} \; 
\tilJ^{\mu\nu}(P^{ab},Q^{fe},Q^{hg}) \;.
\label{gluon_selfD}
\eeq

From Eq.~(\ref{landauQ_PP}),
\beq
(\varPi^{ab,cd}_{\mu \nu})_{\rm gl}(P^{a b}) \HTLapprox
-4 \; g^2 \; \left(
\delta^{ad} \delta^{b c} \; \sum_{e = 1}^{\Nc} \; 
{\tilJ}^{\mu \nu}(P^{ab},Q^{ae},Q^{eb})
- \delta^{ab} \delta^{c d} 
{\tilJ}^{\mu \nu}(P^{ab},Q^{c a},Q^{a c})
\right) \; .
\label{gluon_selfE}
\eeq
Like the quark self-energy, Eq.~(\ref{gluon_self_quark4}), this is traceless,
as it must be, to represent a matrix in $\SUN$.

\begin{figure}
\includegraphics[width=0.75\textwidth]{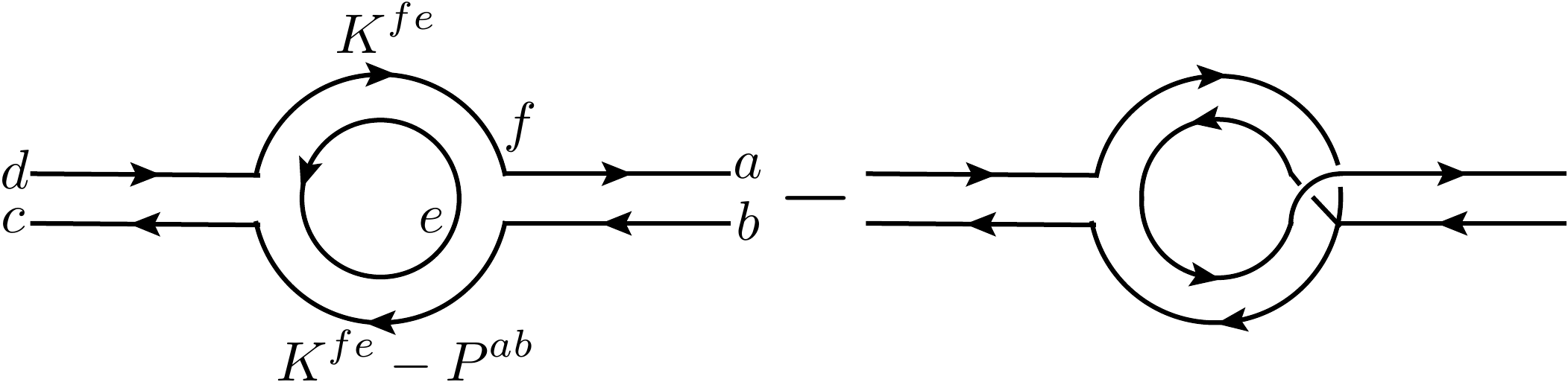}
\caption{One-loop diagram for the gluon loop in the gluon self-energy.}
\label{fig_gluon_loop_gluon_self_energy}
\end{figure}

The color structure is illustrated in the diagram
of Fig.~\ref{fig_gluon_loop_gluon_self_energy}.  There is the planar
diagram, minus a diagram in which the indices are summed over.

In all, the hard thermal loop in the gluon self-energy is the sum of 
Eqs.~(\ref{gluon_self_quark3}) and (\ref{gluon_selfE}).  Each term is
a product of a color dependent factor, times a function of the soft momentum:
\beq
\varPi^{ab,cd}_{\mu \nu}(P^{a b}) \HTLapprox
-{\cal K}^{a b, cd}(Q)\; \delta \varGamma^{\mu \nu}(P^{a b})
- \left(m_{\rm gl}^2\right)^{ab,cd}\!(Q) \; \delta \varPi^{\mu \nu}(P^{a b}) \; .
\label{gluon_selfG}
\eeq

The thermal gluon mass in a background field is
$$
\left(m_{\rm gl}^2\right)^{ab,cd}(Q)
= \frac{g^2 T^2}{6} 
\left( 
\delta^{ad} \delta^{b c} \left[ \; \sum_{e = 1}^{\Nc} \; 
\left( \Acal(Q^{a e})  + \Acal(Q^{e b}) \right)
- \Nf \; \left( \Acal(\aQtl) + \Acal(\bQtl) \right) \right]
\right.
$$
\beq
\left.
- 2\; \delta^{ab} \delta^{c d} \;
\left[ \Acal(Q^{ac}) 
- \frac{\Nf}{\Nc} \left( \Acal(\aQtl) + \Acal(\cQtl)
- \frac{1}{\Nc} \sum_{e = 1}^{\Nc} \Acal(\eQtl) \right)
\right]
\right)
\; .
\label{gluon_selfH}
\eeq
In zero field, $\Acal(0) = 1$, $\Acal(\aQtl) = \Acal(\pi T) = -1/2$,
and
\beq
\left(m_{\rm gl}^2\right)^{ab,cd}(0)
= {\cal P}^{ab, cd} \; \left( \Nc + \frac{\Nf}{2} \right)
\; \frac{g^2 T^2}{3} \; .
\label{gluon_selfI}
\eeq
At $P = 0$, $\varPi^{ab,cd}_{\mu \nu}(0) = \delta^{\mu 0}\delta^{\nu 0} \left(m_{\rm gl}^2\right)^{ab,cd}(0)$,
so Eq. (\ref{gluon_selfI}) is the screening or Debye mass squared for static electric fields at $Q=0$;
Eq. (\ref{gluon_selfH}) is then the generalization to nonzero holonomy.

The new hard thermal loop in the gluon self-energy involves the color
matrix:
\beq
{\cal K}^{a b,cd}(Q)
= \frac{2 i \pi}{3} \,  g^2 T^3
\delta^{ad} \delta^{b c} \left[ \; \sum_{e = 1}^{\Nc} \; 
\left( \nAcal(Q^{a e})  + \nAcal(Q^{e b}) \right)
- \Nf \; \left( \nAcal(\aQtl) + \nAcal(-\bQtl) \right) \right] \;.
\label{gluon_selfJ}
\eeq
The terms proportional to $\delta^{ab}\delta^{cd}$ vanish 
because $\nAcal(Q)$ is odd in $Q$, Eq. (\ref{scalar_k0_C}).
As necessary, the matrix ${\cal K}^{a b,cd}(Q)$ is traceless.

Comparing Eq.~(\ref{gluon_selfJ}) with Eq.~(\ref{total_tadpole}), we find
\beq
{\cal K}^{a b,cd}(Q)=-ig f^{ab,cd,ef}\langle J^{f e}_{0}\rangle \;.
\eeq
 This is natural; the background field induces a color current, which couples to 
 the gluon. The self-energy obeys the Ward-Takahashi identity in the background field:
 \beq
 P_{\mu}^{ab}\varPi_{\mu\nu}^{ab,cd} = - gf^{ab,cd,ef}\langle J^{f e}_{\nu}\rangle \;.
 \eeq

These expressions were computed in Feynman gauge, but the results are
independent of the gauge fixing parameter, $\xi$, Eq.~(\ref{gauge_fixing}).
Except for the gluon self-energy, where the tadpole diagram enters,
the hard thermal loops in any gluon amplitude only involve three gluon
vertices.  This vertex satisfies an Abelian-type Ward
identity, Eq.~(\ref{three_gluon_identity}).  This identity can be used
to show that all hard thermal loops are independent of $\xi$, up to
possibly gauge dependence in terms which are independent of the static
momentum.  However, it is known that at one-loop order that
the potential in a background $Q$ field is independent of $\xi$
\cite{interface1,altes2,altes3}.  Thus, the hard thermal loops are
independent of $\xi$ when $Q \neq 0$, as for $Q=0$.
\section{Conclusions}
\label{sec_conclusions}

In this paper we developed techniques to analyze the real time response
functions for a 't Hooft loop, or $\ZN$ interface.  By introducing
the double line notation in Sec.~\ref{sec_basis}, we are able to analyze
a much more general problem, as is appropriate for the semi-QGP
phase of a gauge theory \cite{hidaka}.  

While the final expressions
which we obtain appear involved, in fact the physics for $Q \neq 0$
is very similar to that for $Q=0$.  For the quark self-energy,
Eq.~(\ref{htl_quark_A}), the hard thermal loop is a thermal quark
mass times a function of momentum.  The function of momentum,
$\delta \Sigma(P)$ in Eq.~(\ref{htl_quark_A1}), is unchanged from
$Q=0$.  What does change with $Q$ is the thermal quark
mass, Eq.~(\ref{htl_quark_A2}).  It is most natural that in the
presence of a nonzero 
background field, that the curvature about the minimum changes with
the background field.  

For the gluon self-energy, Eq.~(\ref{gluon_selfG}), there is a piece
very similar to that for $Q = 0$.  There is the same function of momentum,
$\delta \varPi^{\mu \nu}(P)$ in Eq.~(\ref{landauQ_T}),
as in zero field.  This function is multiplied by
a thermal gluon mass, Eq.~(\ref{gluon_selfH}), 
which is of course $Q$-dependent.

The surprise is that there is a new function in the gluon self-energy,
$\delta\varGamma^{\mu \nu}(P)$ in Eq.~(\ref{landauQ_FB}). 
(Note, however, that this function is
linearly related to $\delta\varPi^{\mu\nu}(P)$, Eq.~(\ref{landauQ_T1})). 
The usual hard thermal loop in the gluon self-energy is $\delta \Pi^{\mu \nu} \sim g^2 T^2$, and is smaller than the new hard thermal loop, $\delta \Gamma^{\mu \nu} \sim g^2T^3/p$.  The general  principle of hard thermal loops, however, is that for soft external momentum, they are as large as the terms at tree level.  This remains valid, since for a non-Abelian gauge field, the propagator in a background field has a new term, $\sim [G^{\mu \nu}, ]$, Eq.~(\ref{field_strength}).

We also note that surely our entire derivation would be much simpler if
we had used kinetic theory in the presence of a nonzero background field.
However, we preferred to use an ordinary perturbative analysis, since 
computing at $Q \neq 0$ is an unfamiliar exercise.  
We do expect that the derivation of a complete
action for all hard thermal loops for nonzero $Q$
would be much simpler with kinetic theory, as it is when $Q=0$.

\acknowledgements
This research of R.D.P. was supported
by the U.S. Department of Energy under
Cooperative Research Agreement No. DE-AC02-98CH10886.
R.D.P. also thanks
the Alexander von Humboldt Foundation for their support.
This research of Y.H. was supported by the Grant-in-Aid for
the Global COE Program ``The Next Generation of Physics, 
Spun from Universality and Emergence'' from the Ministry of 
Education, Culture, Sports, Science and Technology (MEXT) of Japan.
We thank M. Creutz, K. H\"ubner,
F. Karsch, O. Kaczmarek, C. P. Korthals Altes, P. Petreczky, C. Pica, 
R. Venugopalan, and L. Yaffe for discussions.
We especially thank  P.~Cvitanov\'ic
for his detailed comments on Sec.~\ref{sec_basis}.

\appendix*
\section{Z($\Nc$) and U(1) interfaces}
\label{u1_interfaces}

In this appendix we use this opportunity to
make some comments about $\ZN$
\cite{thooft1,thooft2,altes1,interface1,altes2}
and $\U1$ \cite{loop4} interfaces in both Abelian and non-Abelian
gauge theories.

As discussed in the Introduction, Sec.~\ref{introduction},
to define an interface
we pick out one of the three spatial directions, say 
that in the $z$ direction, and consider it separately from the two other
spatial directions, and from that for imaginary time, $\tau$.  
It is also necessary to assume that the length
in the $z$ direction, $L$, is much larger than that for
the two transverse dimensions, of size
$L_t$ \cite{interface1}. 
Lastly, we assume
that both $L$ and $L_t$ are much larger than any physical mass scale,
such as the inverse Debye mass, $\sim 1/(gT)$.

For a pure gauge theory, without dynamical quarks the gauge group is
$\SUN/\ZN$.  Thus we can require that the gauge field at $z = L$ is
a gauge transformation of that at $z =0$.
We take this gauge transformation to be 
a constant element of $\ZN$ \cite{thooft1}:
\beq
\Omega = \exp\left( \frac{2 \pi i k}{\Nc} \right) \; {\bf 1}_N \; .
\eeq
Since this $\Omega$ commutes with all group elements, the gauge field at
$z=L$ is identical to that at
$z=0$.  In going from $z=0$ to $z=L$, though, one winds, in a topologically
nontrivial manner, in the configuration space of gauge potentials
\cite{nair}.  One can show that these boundary conditions are equivalent
to inserting a 't Hooft loop, at $z=L/2$, at the boundary of the two
transverse dimensions \cite{thooft1,altes2}.

There are $k$ distinct transformations possible, where $k=1,2,\ldots(\Nc-1)$.
This generates the cyclic group, $\ZN$, where
$k = \Nc$ is equivalent to the identity.
At tree level, this transformation is implemented by a 
gauge transformation which is linear in $z$, 
\beq
\Omega(z) = \exp\left( 2 \pi i k \; 
\left(\sqrt{2} \; t^{\Nc \Nc} \right)
\; \frac{z}{L} \right) \; .
\eeq
The factor of $\sqrt{2}$ is because of the normalization
for $t^{\Nc \Nc}$ in (\ref{highest_nc}).
This gauge transformation is generated by a constant vector potential
in the imaginary time direction,
$A_0 = (2 \pi i k T/g)
(\sqrt{2}  t^{\Nc \Nc} ) z/L$.
This is a linear ansatz, and is only valid for the tree level action.  
Loop corrections generate a potential for $A_0$, and turn the true solution
into a domain wall, of width $\sim 1/(g T)$
\cite{interface1}.  At present, however, we need not
trouble ourselves with such details.  All that matters is that 
this can be modified to represent a solution
of the quantum equations of motion.

Consider the sum of the following two diagonal generators:
\beq
\lambda_3 =
\sqrt{2} \left( (\Nc - 1) \; t^{\Nc-1, \Nc-1} + t^{\Nc \Nc} \right)
= \sqrt{2(\Nc - 1)(\Nc - 2)}\; \lambda_C^{\Nc-1} \; ,
\label{lambda3}
\eeq
Eq.~(\ref{second_cartan_matrix}).

Consider a gauge transformation generated by this transformation,
$\Omega(z) = \exp(2 \pi i \; \lambda_3 \; z/L) {\bf 1}_N$.  This is 
a ``$\U1$'' transformation, in that it is the same on both ends of the
box, $\Omega(L) = \Omega(0) = {\bf 1}_N$ \cite{loop4}.
This is evident from 
(\ref{lambda3}), where we see that it is a combination of
two $\ZN$ transformations: one of strength 
$\Nc -1 $, along the
$t^{\Nc-1, \Nc-1}$ direction in group space, plus one of unit strength,
along the $t^{\Nc \Nc}$ direction.  The total strength is then $\Nc$,
which for $\ZN$ is equivalent to the identity; i.e., to no winding.

In \cite{loop4}, $\U1$ interfaces were
suggested as a way to perform the semiclassical matching between an
effective and an original theory.  For the pure gauge theory, 
the $\U1$ interfaces are not necessary: we can use the $k$ possible
$\ZN$ interfaces to perform the matching.  
We remark, rather trivially, that if the $\ZN$ interfaces do match,
then so will the $\U1$ interfaces, since by the above, they are just a 
combination of two $\ZN$ interfaces in different directions in group space,
one along $t^{\Nc-1, \Nc-1}$, and the other along $t^{\Nc \Nc}$.

The real use of a $\U1$ interface, however, is for the theory with dynamical
quarks, where the $\ZN$ invariance is broken by the presence of quarks.
Before considering this case, consider a simpler example:
a Higgs model, where a charged scalar field, $\phi$, 
acquires a vacuum expectation value, 
$\langle \phi \rangle$.  Then in going from one end of
the box, to the other, one can 
define a winding number by the number of times that the phase of
$\langle \phi \rangle$ winds around $2 \pi$.
The topology is elementary, just $\pi_1(\U1)\simeq Z$.

Now consider an Abelian gauge theory at nonzero temperature, such as QED,
where the gauge group is unbroken.  
Instead of the expectation value of a Higgs field, we
consider windings of the thermal Wilson line.  The thermal Wilson line is
$ = +1$ at both ends of the box, but it can wind nontrivially as it goes
along the box.  The topology remains $\pi_1(\U1)\simeq Z$.
Thus there are thermal interfaces in QED.  This was first pointed out
by Smilga \cite{smilga}, who computed their properties in the Schwinger
model.  

Thermal $\U1$ interfaces are of interest in the electroweak theory.
They are unlike standard domain walls, in that the interface tension
is strongly dependent upon the temperature.  In the high temperature
phase, where the Higgs field is unbroken,    walls lie
in the $\U1$ for hypercharge.  As the system cools through the electroweak
phase transition, they then rotate into the $\U1$ for electromagnetism.
Since the potential is generated by quantum effects, it vanishes 
exponentially at low temperatures, $\exp(- m_e/T)$, where $m_e$ is the
electron mass.  Thus while $\U1$ domain walls dominate the stress energy
tensor while they exist, unlike standard domain walls,
they naturally vanish at low temperature.  Whether their presence,
during the period in which they dominate the stress energy tensor,
can lead to characteristic cosmological signals is an interesting question.

We return to a $\SUN$ gauge theory.  For each of the $\Nc$
diagonal directions, we can define a $\U1$ winding number.
For example, $\lambda_3$ represents winding once in the first $\Nc - 2$
directions, $- (\Nc - 2)$ times in the $(\Nc-1)^\text{th}$ 
direction, and no winding in the last, $\Nc^\text{th}$ direction.  

Semiclassically, one expands about the configuration along
the $\lambda_3$ direction, in both the original and effective theories.
The configuration is a local minimum of each action, and has
no instabilities under small fluctuations.  Thus such a $\U1$ interface
can be used to match the parameters of the effective theory to the
original theory, as suggested in Ref. \cite{loop4}.

To define such a configuration nonperturbatively, such as on the lattice,
it is necessary to fix the freedom to perform global gauge rotations.
Only after fixing the freedom to change overall gauge rotations can
one define $\U1$ winding numbers for each of the
diagonal directions \cite{loop4}.


\begin{thebibliography}{999}
%
\bibitem{whitepaper}
J.~Adams {\em et.~al.}, \npa{757}{102}{2005}
\href{http://arXiv.org/abs/nucl-ex/0501009}{[arXiv:nucl-ex/0501009]};
K.~Adcox {\em et.~al.}, \ibid{757}{184}{2005}
\href{http://arXiv.org/abs/nucl-ex/0410003}{[arXiv:nucl-ex/0410003]};
I.~Arsene {\em et.~al.}, \ibid{757}{1}{2005}
\href{http://arXiv.org/abs/nucl-ex/0410020}{[arXiv:nucl-ex/0410020]};
B.~B. Back {\em et.~al.}, \ibid{757}{28}{2005}
\href{http://arXiv.org/abs/nucl-ex/0410022}{[arXiv:nucl-ex/0410022]}.
%
\bibitem{strong}
M.~Gyulassy and L.~McLerran,
\npa{750}{30}{2005}
\href{http://arXiv.org/abs/nucl-th/0405013}{[arXiv:nucl-th/0405013]};
A.~Peshier and W.~Cassing,
\prl{94}{172301}{2005}
\href{http://arXiv.org/abs/hep-ph/0502138}{[arXiv:hep-ph/0502138]};
S.~Mrowczynski and M.~H.~Thoma,
\arnps{57}{61}{2007}
\href{http://arXiv.org/abs/nucl-th/0701002}{[arXiv:nucl-th/0701002]};
B.~Muller and J.~L.~Nagle,
\ibid{56}{93}{2006}
\href{http://arXiv.org/abs/nucl-th/0602029}{[arXiv:nucl-th/0602029]};
E. V. Shuryak,
to appear in Prog. Part. Nucl. Phys.
\href{http://arXiv.org/abs/0807.3033}{[arXiv:0807.3033]};
R.~D.~Pisarski, PoS {\bf LATTICE2008}, 016 (2008);
\href{http://arXiv.org/abs/0810.4585}{[arXiv:0810.4585]};
%
U.~W.~Heinz,
\href{http://arXiv.org/abs/0901.4355}{[arXiv:0901.4355]}.
%
\bibitem{thooft1}
G.~'t Hooft,
\npb{138}{1}{1978}.
%
\bibitem{thooft2}
G.~'t Hooft,
\npb{153}{141}{1979}.
%
\bibitem{altes1}
C.~P. ~Korthals-Altes, A.~Kovner and M.~A.~Stephanov,
\plb{469}{205}{1999}
\href{http://arXiv.org/abs/hep-ph/9909516}{[arXiv:hep-ph/9909516]};
C.~P.~Korthals Altes and A.~Kovner,
\prd{62}{096008}{2000}
\href{http://arXiv.org/abs/hep-ph/0004052}{[arXiv:hep-ph/0004052]}.
%
\bibitem{interface1}
T.~Bhattacharya, A.~Gocksch, C.~P.~Korthals Altes and R.~D.~Pisarski,
\prl{66}{998}{1991};
\npb{383}{497}{1992}
\href{http://arXiv.org/abs/hep-ph/9205231}{[arXiv:hep-ph/9205231]},
and references therein.
%
\bibitem{altes2}
C.~P.~Korthals Altes,
\npb{420}{637}{1994}
\href{http://arXiv.org/abs/hep-th/9310195}{[arXiv:hep-th/9310195]};
P.~Giovannangeli and C.~P.~Korthals Altes,
\ibid{608}{203}{2001}
\href{http://arXiv.org/abs/hep-ph/0102022}{[arXiv:hep-ph/0102022]};
\ibid{721}{1}{2005}
\href{http://arXiv.org/abs/hep-ph/0212298}{[arXiv:hep-ph/0212298]};
\ibid{721}{25}{2005}
\href{http://arXiv.org/abs/hep-ph/0412322}{[arXiv:hep-ph/0412322]}.
%
\bibitem{altes3}
C. P. ~Korthals Altes, 
\href{http://arXiv.org/abs/0904.3117}{[arXiv:0904.3117]},
and private communication.
%
\bibitem{linear_ZN}
A.~Vuorinen and L.G.~Yaffe,
\prd{74}{025011}{2006}
\href{http://arXiv.org/abs/hep-ph/0604100}{[arXiv:hep-ph/0604100]};
A.~Kurkela,
\prd{76}{094507}{2007}
\href{http://arXiv.org/abs/0704.1416}{[arXiv:0704.1416]};
Ph.~de Forcrand, A.~Kurkela and A.~Vuorinen,
\prd{77}{125014}{2008}
\href{http://arXiv.org/abs/0801.1566}{[arXiv:0801.1566]};
C. P. ~Korthals Altes, 
\href{http://arXiv.org/abs/0810.3325}{[arXiv:0810.3325]}.
%
\bibitem{interface_lattice}
C.~P.~Korthals Altes, A.~Michels, M.~A.~Stephanov and M.~Teper,
\prd{55}{1047}{1997}
\href{http://arXiv.org/abs/hep-lat/9606021}{[arXiv:hep-lat/9606021]};
P.~de Forcrand, M.~D'Elia and M.~Pepe,
\prl{86}{1438}{2001}
\href{http://arXiv.org/abs/hep-lat/0007034}{[arXiv:hep-lat/0007034]};
P.~de Forcrand and L.~von Smekal,
\prd{66}{011504(R)}{2002}
\href{http://arXiv.org/abs/hep-lat/0107018}{[arXiv:hep-lat/0107018]};
F.~Bursa and M.~Teper,
\jhep{0508}{060}{2005}
\href{http://arXiv.org/abs/hep-lat/0505025}{[arXiv:hep-lat/0505025]};
P.~de Forcrand and D.~Noth,
\prd{72}{114501}{2005}
\href{http://arXiv.org/abs/hep-lat/0506005}{[arXiv:hep-lat/0506005]};
Ph.~de Forcrand, C.~P.~Korthals Altes and O.~Philipsen,
\npb{742}{124}{2006}
\href{http://arXiv.org/abs/hep-ph/0510140}{[arXiv:hep-ph/0510140]}.
%
\bibitem{real_time_lattice}
A.~Bazavov, B.~A.~Berg and A.~Velytsky,
\prd{74}{014501}{2006}
\href{http://arXiv.org/abs/hep-lat/0605001}{[arXiv:hep-lat/0605001]};
A.~Bazavov, B.~A.~Berg and A.~Dumitru,
\ibid{78}{034024}{2008}
\href{http://arXiv.org/abs/0805.0784}{[arXiv:0805.0784]}.
%
\bibitem{earlyA}
D.~J.~Gross, R.~D.~Pisarski and L.~G.~Yaffe,
\rmp{53}{43}{1981}.
%
\bibitem{earlyB}
N.~Weiss,
\prd{24}{475}{1981};
\ibid{25}{2667}{1982}.
%
\bibitem{imag_chem_pot}
A.~Roberge and N.~Weiss,
\npb{275}{734}{1986};
A.~Gocksch and R.~D.~Pisarski,
\ibid{402}{657}{1993}
\href{http://arXiv.org/abs/hep-ph/9302233}{[arXiv:hep-ph/9302233]}.
%
\bibitem{smilga}
A.~V.~Smilga,
\anp{234}{1}{1994};
\phr{291}{1}{1997}
\href{http://arXiv.org/abs/hep-ph/9612347}{[arXiv:hep-ph/9612347]}.
%
\bibitem{background}
G.~'t Hooft and M.~J.~G.~Veltman,
\npb{44}{189}{1972};
L.~F.~Abbott,
\ibid{185}{189}{1981}.
%
%
\bibitem{resurgent}
U.~D.~Jentschura and J.~Zinn-Justin,
\plb{596}{138}{2004}
\href{http://arXiv.org/abs/hep-ph/0405279}{[arXiv:hep-ph/0405279]}.
%
\bibitem{braaten}
E.~Braaten and A.~Nieto,
\prd{53}{3421}{1996}
\href{http://arXiv.org/abs/hep-ph/9510408}{[arXiv:hep-ph/9510408]}.
%
\bibitem{perturbative}
P.~Arnold and C.~Zhai,
\prd{50}{7603}{1994}
\href{http://arXiv.org/abs/hep-ph/9408276}{[arXiv:hep-ph/9408276]};
\ibid{51}{1906}{1995}
\href{http://arXiv.org/abs/hep-ph/9410360}{[arXiv:hep-ph/9410360]};
C.~Zhai and B.~Kastening,
\ibid{52}{7232}{1995}
\href{http://arXiv.org/abs/hep-ph/9507380}{[arXiv:hep-ph/9507380]}.
%
\bibitem{thermal_review}
U.~Kraemmer and A.~Rebhan,
\rpp{67}{351}{2004}
\href{http://arXiv.org/abs/hep-ph/0310337}{[arXiv:hep-ph/0310337]};
J.~O.~Andersen and M.~Strickland,
\anp{317}{281}{2005}
\href{http://arXiv.org/abs/hep-ph/0404164}{[arXiv:hep-ph/0404164]}.
%
\bibitem{resum}
K.~Kajantie, M.~Laine, K.~Rummukainen and Y.~Schr\"oder,
\prl{86}{10}{2001}
\href{http://arXiv.org/abs/hep-ph/0007109}{[arXiv:hep-ph/0007109]};
\prd{67}{105008}{2003}
\href{http://arXiv.org/abs/hep-ph/0211321}{[arXiv:hep-ph/0211321]};
A.~Hietanen, K.~Kajantie, M.~Laine, K.~Rummukainen and Y.~Schr\"oder,
\jhep{0501}{013}{2005}
\href{http://arXiv.org/abs/hep-lat/0412008}{[arXiv:hep-lat/0412008]};
F.~Di Renzo, M.~Laine, V.~Miccio, Y.~Schr\"oder and 
C.~Torrero, \jhep{0607}{026}{2006}
\href{http://arXiv.org/abs/hep-ph/0605042}{[arXiv:hep-ph/0605042]};
F.~Di Renzo, M.~Laine, Y.~Schr\"oder and C.~Torrero,
\ibid{0809}{061}{2008}
\href{http://arXiv.org/abs/0808.0557}{[arXiv:0808.0557]};
A.~Hietanen, K.~Kajantie, M.~Laine, K.~Rummukainen and Y.~Schroder,
\prd{79}{045018}{2009}
\href{http://arXiv.org/abs/0811.4664}{[arXiv:0811.4664]}.
%
\bibitem{pert_coupling}
M.~Laine and Y.~Schr\"oder, \jhep{0503}{067}{2005}
\href{http://arXiv.org/abs/hep-ph/0503061}{[arXiv:hep-ph/0503061]};
\href{http://arXiv.org/abs/hep-lat/0509104}{[arXiv:hep-lat/0509104]};
\prd{73}{085009}{2006}
\href{http://arXiv.org/abs/hep-ph/0603048}{[arXiv:hep-ph/0603048]};
P.~Giovannangeli,
\npb{738}{23}{2006}
\href{http://arXiv.org/abs/hep-ph/0506318}{[arXiv:hep-ph/0506318]}.
%
\bibitem{htlA}
R.~D.~Pisarski,
\prl{63}{1129}{1989};
E.~Braaten and R.~D.~Pisarski,
\ibid{64}{1338}{1990}.
%
\bibitem{htlB}
E.~Braaten and R.~D.~Pisarski,
\npb{337}{569}{1990}.
%
%
\bibitem{lebellac}
M. Le Bellac, {\it Thermal Field Theory} (Cambridge University Press,
Cambridge, 2000), and references therein.
%
\bibitem{loop1}
R.~D.~Pisarski,
\prd{62}{111501(R)}{2000}
\href{http://arXiv.org/abs/hep-ph/0006205}{[arXiv:hep-ph/0006205]};
A.~Dumitru and R.~D.~Pisarski,
\plb{504}{282}{2001}
\href{http://arXiv.org/abs/hep-ph/0010083}{[arXiv:hep-ph/0010083]};
\ibid{525}{95}{2002}
\href{http://arXiv.org/abs/hep-ph/0106176}{[arXiv:hep-ph/0106176]};
\prd{66}{096003}{2002}
\href{http://arXiv.org/abs/hep-ph/0204223}{[arXiv:hep-ph/0204223]};
O.~Scavenius, A.~Dumitru and J.~T.~Lenaghan,
\prc{66}{034903}{2002}
\href{http://arXiv.org/abs/hep-ph/0201079}{[arXiv:hep-ph/0201079]}.
%
\bibitem{loop2}
A.~Dumitru, Y.~Hatta, J.~Lenaghan, K.~Orginos and R.~D.~Pisarski,
\prd{70}{034511}{2004}
\href{http://arXiv.org/abs/hep-th/0311223}{[arXiv:hep-th/0311223]}.
%
\bibitem{loop3}
A.~Dumitru, J.~Lenaghan and R.~D.~Pisarski,
\prd{71}{074004}{2005}
\href{http://arXiv.org/abs/hep-ph/0410294}{[arXiv:hep-ph/0410294]};
M. Oswald and R. D. Pisarski, 
\ibid{74}{045029}{2006}
\href{http://arXiv.org/abs/hep-ph/0512245}{[arXiv:hep-ph/0512245]}.
%
\bibitem{loop4}
R.~D.~Pisarski,
\prd{74}{121703(R)}{2006}
\href{http://arXiv.org/abs/hep-ph/0608242}{[arXiv:hep-ph/0608242]}.
%
\bibitem{lattice_effective}
L.~Dittmann, T.~Heinzl and A.~Wipf,
\jhep{0406}{005}{2004}
\href{http://arXiv.org/abs/hep-lat/0306032}{[arXiv:hep-lat/0306032]};
T.~Heinzl, T.~Kaestner and A.~Wipf,
\prd{72}{065005}{2005}
\href{http://arXiv.org/abs/hep-lat/0502013}{[arXiv:hep-lat/0502013]};
C.~Wozar, T.~Kaestner, A.~Wipf, T.~Heinzl and B.~Pozsgay,
\ibid{74}{114501}{2006}
\href{http://arXiv.org/abs/hep-lat/0605012}{[arXiv:hep-lat/0605012]};
C.~Wozar, T.~Kaestner, A.~Wipf and T.~Heinzl,
\ibid{76}{085004}{2007}
\href{http://arXiv.org/abs/0704.2570}{[arXiv:0704.2570]};
A.~Dumitru and D.~Smith,
\ibid{77}{094022}{2008}
\href{http://arXiv.org/abs/0711.0868}{[arXiv:0711.0868]};
A.~Velytsky,
\ibid{78}{034505}{2008}
\href{http://arXiv.org/abs/0805.4450}{[arXiv:0805.4450]};
C.~Wozar, T.~Kastner, B.~H.~Wellegehausen, A.~Wipf and T.~Heinzl
\href{http://arXiv.org/abs/0808.4046}{[arXiv:0808.4046]}.
%
\bibitem{pnjl}
K.~Fukushima,
\prd{68}{045004}{2003}
\href{http://arXiv.org/abs/hep-ph/0303225}{[arXiv:hep-ph/0303225]};
\plb{591}{277}{2004}
\href{http://arXiv.org/abs/hep-ph/0310121}{[arXiv:hep-ph/0310121]};
Y.~Hatta and K.~Fukushima,
\prd{69}{097502}{2004}
\href{http://arXiv.org/abs/hep-ph/0307068}{[arXiv:hep-ph/0307068]};
A.~Dumitru, R.~D.~Pisarski and D.~Zschiesche,
\ibid{72}{065008}{2005}
\href{http://arXiv.org/abs/hep-ph/0505256}{[arXiv:hep-ph/0505256]};
C.~Ratti, M.~A.~Thaler and W.~Weise,
\ibid{73}{014019}{2006}
\href{http://arXiv.org/abs/hep-ph/0506234}{[arXiv:hep-ph/0506234]};
S.~K.~Ghosh, T.~K.~Mukherjee, M.~G.~Mustafa and R.~Ray,
\ibid{73}{114007}{2006}
\href{http://arXiv.org/abs/hep-ph/0603050}{[arXiv:hep-ph/0603050]};
E.~Megias, E.~Ruiz Arriola and L.~L.~Salcedo,
\ibid{74}{065005}{2006}
\href{http://arXiv.org/abs/hep-ph/0412308}{[arXiv:hep-ph/0412308]};
\jhep{0601}{073}{2006}
\href{http://arXiv.org/abs/hep-ph/0505215}{[arXiv:hep-ph/0505215]};
\prd{74}{114014}{2006}
\href{http://arXiv.org/abs/hep-ph/0607338}{[arXiv:hep-ph/0607338]};
H.~Hansen, W.~M.~Alberico, A.~Beraudo, A.~Molinari, M.~Nardi and C.~Ratti,
\ibid{75}{065004}{2007}
\href{http://arXiv.org/abs/hep-ph/0609116}{[arXiv:hep-ph/0609116]};
S.~Mukherjee, M.~G.~Mustafa and R.~Ray,
\ibid{75}{094015}{2007}
\href{http://arXiv.org/abs/hep-ph/0609249}{[arXiv:hep-ph/0609249]};
S.~Roessner, C.~Ratti and W.~Weise,
\ibid{75}{034007}{2007}
\href{http://arXiv.org/abs/hep-ph/0609281}{[arXiv:hep-ph/0609281]};
K.~Fukushima and Y.~Hidaka,
\ibid{75}{036002}{2007}
\href{http://arXiv.org/abs/hep-ph/0610323}{[arXiv:hep-ph/0610323]};
C.~Sasaki, B.~Friman and K.~Redlich,
\ibid{75}{074013}{2007}
\href{http://arXiv.org/abs/hep-ph/0611147}{[arXiv:hep-ph/0611147]};
E.~Megias, E.~Ruiz Arriola and L.~L.~Salcedo,
\ibid{75}{105019}{2007}
\href{http://arXiv.org/abs/hep-ph/0702055}{[arXiv:hep-ph/0702055]};
S.~K.~Ghosh, T.~K.~Mukherjee, M.~G.~Mustafa and R.~Ray,
\ibid{77}{094024}{2008}
\href{http://arXiv.org/abs/0710.2790}{[arXiv:0710.2790]};
S.~Roessner, T.~Hell, C.~Ratti and W.~Weise,
\npa{814}{118}{2008}
\href{http://arXiv.org/abs/0712.3152}{[arXiv:0712.3152]};
P.~Costa, C.~A.~de Sousa, M.~C.~Ruivo and H.~Hansen,
Europhys. Lett. 86, 31001 (2009)
\href{http://arXiv.org/abs/0801.3616}{[arXiv:0801.3616]};
Y.~Sakai, K.~Kashiwa, H.~Kouno and M.~Yahiro,
\prd{78}{036001}{2008}
\href{http://arXiv.org/abs/0803.1902}{[arXiv:0803.1902]};
K.~Fukushima,
\ibid{77}{114028}{2008} [Erratum-\ibid{78}{039902}{2008}]
\href{http://arXiv.org/abs/0803.3318}{[arXiv:0803.3318]};
\href{http://arXiv.org/abs/0809.3080}{[arXiv:0809.3080]};
P.~Costa, M.~C.~Ruivo, C.~A.~de Sousa, H.~Hansen and W.~M.~Alberico,
\prd{79}{116003}{2009}
\href{http://arXiv.org/abs/0807.2134}{[arXiv:0807.2134]};
K.~Dusling, C.~Ratti and I.~Zahed,
\prd{79}{034027}{2009}
\href{http://arXiv.org/abs/0807.2879}{[arXiv:0807.2879]};
T.~Hell, S.~Roessner, M.~Cristoforetti and W.~Weise,
\prd{79}{014022}{2009}
\href{http://arXiv.org/abs/0810.1099}{[arXiv:0810.1099]};
%
E.~Megias, E.~Ruiz Arriola and L.~L.~Salcedo,
\prd{80}{056005}{2009}
\href{http://arXiv.org/abs/0903.1060}{[arXiv:0903.1060]}.
\bibitem{hidaka}
Y.~Hidaka and R.~D.~Pisarski
\prd{78}{071501(R)}{2008}
\href{http://arXiv.org/abs/0803.0453}{[arXiv:0803.0453]}.
%
\bibitem{small_sphere1}
B.~Sundborg,
\npb{573}{349}{2000}
\href{http://arXiv.org/abs/hep-th/9908001}{[arXiv:hep-th/9908001]};
O.~Aharony, J.~Marsano, S.~Minwalla, K.~Papadodimas and M.~Van Raamsdonk,
\atmp{8}{603}{2004}
\href{http://arXiv.org/abs/hep-th/0310285}{[arXiv:hep-th/0310285]};
\prd{71}{125018}{2005}           
\href{http://arXiv.org/abs/hep-th/0502149}{[arXiv:hep-th/0502149]};       
H.~J.~Schnitzer,
\npb{695}{267}{2004}
\href{http://arXiv.org/abs/hep-th/0402219}{[arXiv:hep-th/0402219]};
O.~Aharony, J.~Marsano, S.~Minwalla and T.~Wiseman,
\cqg{21}{5169}{2004}
\href{http://arXiv.org/abs/hep-th/0406210}{[arXiv:hep-th/0406210]};
L.~Alvarez-Gaume, C.~Gomez, H.~Liu and S.~Wadia,
\prd{71}{124023}{2005}
\href{http://arXiv.org/abs/hep-th/0502227}{[arXiv:hep-th/0502227]};
S.~A.~Hartnoll and S.~P.~Kumar,
\ibid{76}{026005}{2007}
\href{http://arXiv.org/abs/hep-th/0610103}{[arXiv:hep-th/0610103]};
U.~Gursoy, S.~A.~Hartnoll, T.~J.~Hollowood and S.~P.~Kumar,
\jhep{0711}{020}{2007}
\href{http://arXiv.org/abs/hep-th/0703100}{[arXiv:hep-th/0703100]}.
%
\bibitem{small_sphere2}
O.~Aharony, J.~Marsano, S.~Minwalla, K.~Papadodimas and M.~Van Raamsdonk,
\prd{71}{125018}{2005}           
\href{http://arXiv.org/abs/hep-th/0502149}{[arXiv:hep-th/0502149]}.
%
\bibitem{ren_loop1}
J.-L.~Gervais and A.~Neveu, \npb{163}{189}{1980};
A.~M.~Polyakov, \ibid{164}{171}{1980};
V.~S.~Dotsenko and S.~N.~Vergeles, \ibid{169}{527}{1980};
I.~Y.~Arefeva, \plb{93}{347}{1980};
R. A. Brandt, F. Neri, and M. Sato, \prd{24}{879}{1981};
R. A. Brandt, A. Gocksch, F. Neri, and M. Sato, \ibid{26}{3611}{1982}.
%
\bibitem{ren_loop2}
E.~Gava and R.~Jengo,
\plb{105}{285}{1981}.
%
\bibitem{renloop_lattice1}
O.~Kaczmarek, F.~Karsch, P.~Petreczky and F.~Zantow,
\plb{543}{41}{2002}
\href{http://arXiv.org/abs/hep-lat/0207002}{[arXiv:hep-lat/0207002]};
P.~Petreczky and K.~Petrov,
\prd{70}{054503}{2004}
\href{http://arXiv.org/abs/hep-lat/0405009}{[arXiv:hep-lat/0405009]};
O.~Kaczmarek, F.~Karsch, F.~Zantow and P.~Petreczky,
\ibid{70}{074505}{2004}
[Erratum-\ibid{72}{059903}{2005}]
\href{http://arXiv.org/abs/hep-lat/0406036}{[arXiv:hep-lat/0406036]};
M.~Doring, S.~Ejiri, O.~Kaczmarek, F.~Karsch and E.~Laermann,
\epjc{46}{179}{2006}
\href{http://arXiv.org/abs/hep-lat/0509001}{[arXiv:hep-lat/0509001]};
K.~Hubner, F.~Karsch, O.~Kaczmarek and O.~Vogt,
\prd{77}{074504}{2008}
\href{http://arXiv.org/abs/0710.5147}{[arXiv:0710.5147]}.
%
\bibitem{renloop_lattice2}
Y.~Aoki, Z.~Fodor, S.~D.~Katz and K.~K.~Szabo,
\plb{643}{46}{2006}
\href{http://arXiv.org/abs/hep-lat/0609068}{[arXiv:hep-lat/0609068]}.
%
\bibitem{renloop_lattice3}
M.~Cheng {\it et al.},
\prd{77}{014511}{2008}
\href{http://arXiv.org/abs/0710.0354}{[arXiv:0710.0354]}.
%
\bibitem{ghk}
S.~Gupta, K.~H\"ubner and O.~Kaczmarek,
\prd{77}{034503}{2008}
\href{http://arXiv.org/abs/0711.2251}{[arXiv:0711.2251]}.
%
\bibitem{furuuchi}
K.~Furuuchi,
\prd{73}{046004}{2006}
\href{http://arXiv.org/abs/hep-th/0510056}{[arXiv:hep-th/0510056]};
\href{http://arXiv.org/abs/hep-th/0608181}{[arXiv:hep-th/0608181]}.
%
%
\bibitem{thooft_planar}
G. 't Hooft,
\npb{72}{461}{1974};
``Planar Diagram Field Theories'', in
{\it Progress In Gauge Field Theory. Proceedings, NATO Advanced Study Institute}, 
G.~'t Hooft, A.~Jaffe, H.~Lehmann, P.~K.~Mitter, I.~M.~Singer and R.~Stora, editors (Plenum Press, NY, 1984).
%
\bibitem{cvitanovic}
P.~Cvitanov\'ic,
\prd{14}{1536}{1976};
{\it Group Theory: Birdtracks, Lie's, and Exceptional Groups}
(Princeton University Press, Princeton, 2008)
\href{http://birdtracks.eu/}{[birdtracks.eu]}.
%
\bibitem{nair}
V. P. Nair, ``Quantum Field Theory'' (Springer, Berlin, 2005),
especially Secs. 15.5 and 20.4.1.
%
\end{thebibliography}
\end{document}